\documentclass[]{aastex62}

\defcitealias{Rodriguez:2022}{R22}
\defcitealias{Mroz2020B}{M20}

\usepackage{scalerel}
\usepackage{tikz}
\usetikzlibrary{svg.path}

\newcommand{\orcid}[1]{\href{https://orcid.org/#1}{\textcolor[HTML]{A6CE39}{\aiOrcid}}}

\definecolor{orcidlogocol}{HTML}{A6CE39}
\tikzset{
    orcidlogo/.pic={
        \fill[orcidlogocol] svg{M256,128c0,70.7-57.3,128-128,128C57.3,256,0,198.7,0,128C0,57.3,57.3,0,128,0C198.7,0,256,57.3,256,128z};
        \fill[white] svg{M86.3,186.2H70.9V79.1h15.4v48.4V186.2z}
        svg{M108.9,79.1h41.6c39.6,0,57,28.3,57,53.6c0,27.5-21.5,53.6-56.8,53.6h-41.8V79.1z M124.3,172.4h24.5c34.9,0,42.9-26.5,42.9-39.7c0-21.5-13.7-39.7-43.7-39.7h-23.7V172.4z}
        svg{M88.7,56.8c0,5.5-4.5,10.1-10.1,10.1c-5.6,0-10.1-4.6-10.1-10.1c0-5.6,4.5-10.1,10.1-10.1C84.2,46.7,88.7,51.3,88.7,56.8z};
    }
}

\newcommand\orcidicon[1]{\href{https://orcid.org/#1}{\mbox{\scalerel*{
                \begin{tikzpicture}[yscale=-1,transform shape]
                \pic{orcidlogo};
                \end{tikzpicture}
            }{|}}}}

\usepackage{amssymb}
\usepackage{amsmath}
\usepackage{graphicx}
\usepackage{dcolumn}
\usepackage{hyperref}
\usepackage{longtable}
\hypersetup{
    colorlinks=true,
    citecolor=blue,
    linkcolor=blue,
    urlcolor=blue,
}
\usepackage[utf8]{inputenc}

\renewcommand{\edit}[1]{{\textcolor{black}{#1}}}

\shorttitle{60 Microlensing Events in ZTF One} 
\shortauthors{Medford \& Abrams et al.}

\begin{document}

\setlength{\abovedisplayskip}{5pt}
\setlength{\belowdisplayskip}{5pt}

\title{60 Microlensing Events from the Three Years of Zwicky Transient Facility Phase One}

\author{Michael S. Medford \orcidicon{0000-0002-7226-0659}}
\affiliation{University of California, Berkeley, Department of Astronomy, Berkeley, CA 94720}
\affiliation{Lawrence Berkeley National Laboratory, 1 Cyclotron Rd., Berkeley, CA 94720}

\author{Natasha S. Abrams \orcidicon{0000-0002-0287-3783}}
\affiliation{University of California, Berkeley, Department of Astronomy, Berkeley, CA 94720}

\author{Jessica R. Lu \orcidicon{0000-0001-9611-0009}}
\affiliation{University of California, Berkeley, Department of Astronomy, Berkeley, CA 94720}

\author{Peter Nugent \orcidicon{0000-0002-3389-0586}}
\affiliation{Lawrence Berkeley National Laboratory, 1 Cyclotron Rd., Berkeley, CA 94720}

\author{Casey Y. Lam \orcidicon{0000-0002-6406-1924}}
\affiliation{University of California, Berkeley, Department of Astronomy, Berkeley, CA 94720}

\begin{abstract}

Microlensing events have historically been discovered throughout the Galactic bulge and plane by surveys designed solely for that purpose.
We conduct the first multi-year search for microlensing events on the Zwicky Transient Facility (ZTF), an all-sky optical synoptic survey that observes the entire visible Northern sky every few nights. 
We discover 60 high quality microlensing events in the three years of ZTF-I using the bulk lightcurves in the ZTF Public Data Release 5.
19 of our events are found outside of the Galactic plane ($|b| \geq 15^\circ$), nearly doubling the number of previously discovered events in the stellar halo from surveys pointed toward the Magellanic Clouds and the Andromeda Galaxy.
We also record 1,558 ongoing candidate events as potential microlensing that can continue to be observed by ZTF-II for identification.
The scalable and computationally efficient methods developed in this work can be applied to future synoptic surveys, such as the Vera C. Rubin Observatory's Legacy Survey of Space and Time and the Nancy Grace Roman Space Telescope, as they attempt to find microlensing events in even larger and deeper datasets.

\end{abstract}


\section{Introduction} \label{sec:intro}

\citet{Einstein1936LENS-LIKEFIELD} first derived that objects with mass could bend the light of a luminous star on its way to an observer and produce multiple images of the background star.
This phenomenon is called gravitational microlensing because the individual images are separated by microarcseconds and are therefore unresolvable to any instrument.
The observer can only measure a lightcurve that includes both the constant luminosity of the lens and the apparent increase in the brightness of the background star \citep{Refsdal1964}.
Microlensing events will last as long as there is an apparent alignment, occurring on timescales ranging from days to years for one star magnifying another within the Milky Way \edit{G}alaxy \citep{Paczynski1986}.

\edit{The amplification of the luminous source is maximized at the time ($t_0$) when the projected separation between the source and lens on the sky is minimized. The degree of amplification is set by the impact parameter ($u_0$), which is the minimum separation divided by the characteristic interaction cross-section or Einstein radius, 
\begin{equation}
\theta_E = 2.85 \;\textrm{mas} \left(\frac{M}{M_\odot}\right)^{\edit{1/2}} \left(\frac{\pi_{rel}}{\textrm{1 mas}}\right)^{\edit{1/2}}
\end{equation}
where $\pi_{rel} = \pi_L - \pi_S$ which is the difference in parallax between the lens and source. 
The duration of this amplification, named the Einstein crossing time ($t_E$), is the time over which $\theta_E$ is traversed by the relative proper motion between the lens and source ($\mu_\text{rel}$),}
\begin{equation}
    t_E = \theta_E / \mu_\text{rel}\ .
\end{equation}
For more thorough \edit{definitions} of these parameters we refer the reader to \citet{Gould1992}.

Microlensing can be distinguished from other astrophysical transients due to several unique characteristics.
The amplification of a source is ideally followed by a decrease in the source's brightness that is symmetric across the point of maximum amplification when a constant velocity for the lens and source is assumed.
However, this symmetry is complicated by the non-uniform motion of the Earth as it orbits around the sun \citep{Gould1992}, which produces a \edit{quasi}-annual variation to the source amplification \edit{known as the microlensing parallax ($\pi_E$),}
\begin{eqnarray}
    \pi_E &=& \pi_\text{rel} / \theta_E\ .
\end{eqnarray}
Either the symmetry of the observable lightcurve when there is low parallax or the presence of an annual variation to the amplification where there is high parallax is evidence of microlensing.

The amplification of the background star is achromatic as the geometric effects of bending space-time occur equally for all wavelengths of light.
However, the constant presence of the lens light in the lightcurve will produce a chromatic effect called blending \citep{DiStefano1995}.
Imagine a source star that \edit{only emits flux at a red wavelength and a lens star of equal flux that only emits at a blue wavelength.
Before (and after) the microlensing event, the brightness measured in corresponding red and blue filters could be equal to each other, but during the event the red filter would increase in brightness while the blue filter measured a constant flux, leading to a chromatic effect.}
This is further complicated by the presence of multiple neighbor stars not close enough to be involved in the microlensing phenomenon but within the point-spread function of the instrument.
These stars contaminate the lightcurve and also contribute additional unamplified light that causes chromatic blending.
Together these effects produce a chromatic signal that can be modeled as a combination of unamplified and amplified light.
The fraction of light that originates from the source (as compared to the light from the unamplified lens and neighbors) is given the name source-flux-fraction or $b_\text{sff}$.

The probability of two stars crossing the same line of sight is proportional to the apparent stellar density.
This has motivated most microlensing surveys to look for events toward the Galactic bulge \citep{Sumi2013, Udalski2015, Navarro2017, Kim_2018, Mrz2019}, as well as the Magellanic \edit{C}louds \citep{Alcock2000, Tisserand2007, Wyrzykowski2011a, Wyrzykowski2011b} and M31 \citep{Calchi_Novati_2014, Niikura_2019}.
These surveys discover hundreds to thousands of microlensing events per year.
There have been fewer microlensing surveys dedicated to looking for these events throughout the Galactic plane.
The Exp\'erience pour la Recherche d'Objets Sombres (EROS) spiral arm surveys \citep{Rahal2009} observed 12.9 million stars across four directions in the Galactic plane and discovered 27 microlensing events over seven years between 1996 and 2002.
The Optical Gravitational Lensing Experiment (OGLE; \citet{Mroz2020A}) discovered 630 events in 3000 square degrees of Galactic plane fields in the 7 years between 2013 and 2019.
They measured a threefold increase in the average Einstein crossing time for Galactic plane events as compared to the Galactic bulge and an asymmetric optical depth that they interpret as evidence of the Galactic warp.
\citet{Mroz2020B} found 30 microlensing candidates in the first year of Zwicky Transient Facility (ZTF) operations.

While the number of events discovered thus far are much smaller than in Galactic bulge fields, simulations predict that there are many more events left to discover in large synoptic surveys.
\citet{Sajadian_2019} predict that the Vera Rubin Observatory Legacy Survey of Space and Time (LSST) \edit{could} discover anywhere between 7900 and 34000 microlensing events over ten years of operation depending on its observing strategy.
\citet{Medford2020b} estimate that ZTF would discover $\sim$1100 detectable Galactic plane microlensing events in its first three years of operation, with $\sim$500 of these events occurring outside of the Galactic bulge ($\ell \geq 10^\circ$).
Microlensing events throughout the Galactic plane can yield interesting information about Galactic structure and stellar evolution that we cannot learn from only looking toward the Galactic bulge.

Gravitational lenses only need mass but not luminosity in order to create a microlensing event, thereby making microlensing the only method available for detecting a particularly interesting non-luminous lens object: isolated black holes.
Inspired by the suggestion \edit{of} \citet{Paczynski1986} that places constraints on the amount of dark matter found in MAssive Halo Compact Objects (MACHOs) \edit{that} could be measured by microlensing, there had been ongoing efforts to find these MACHOs, including black holes, as gravitational lenses for many years.
The MACHO \citep{Alcock2000} and EROS \citep{Afonso2003} projects calculated these upper limits after several years of observations toward the Galactic bulge and Magellanic \edit{C}louds.
While black hole candidates have been proposed from individual microlensing lightcurves \citep{Bennett2002, Wyrzykowski2016}, there is a degeneracy between the mass of the lens and the distances to the source and lens that cannot be broken through photometry alone.
Essentially, photometry cannot distinguish between a massive lens that is relatively distant and a less massive lens that is closer to the observer.
\citet{Lu2016} outlined how direct measurement of the apparent shift in the centroid of the unresolved source and lens, or astrometric microlensing, can be used to break this degeneracy and confirm black hole microlensing candidates.
This shift is extremely small, on the order of milliarcseconds, and therefore requires high resolution measurements on an adaptive optics system such as Keck AO.
Several candidates have been followed up using this technique
\edit{\citep{Lu2016} and one possible black hole has been detected \citep{Lam2022,Sahu2022, Mroz2022}.
}
Only a few can be rigorously followed up astrometrically and the astrometric signal of events found toward the bulge is at the edge of current observational capabilities.
Simulations indicate that black hole candidates can be ideally selected from a list of microlensing events by searching for those with larger Einstein crossing time and smaller microlensing parallaxes \citep{Lam2020}.
Furthermore, simulations indicate that events in the Galactic plane have larger microlensing parallaxes and astrometric shifts overall that can make finding black hole candidates amongst these surveys easier than surveys pointed toward the Galactic bulge \citep{Medford2020b}.

In this paper, we will conduct a search for microlensing events in the Zwicky Transient Facility dataset that is optimized to find these black hole candidates, increasing the number of candidates that can be astrometrically followed up for lens mass confirmation. \edit{Though the approach is mildly optimized for long duration events, the pipeline is equipped to find microlensing events with a range of durations.}
A previous version of this approach appeared in the thesis \citet{Medford2021}.
In Section \ref{sec:telescope} we discuss the Zwicky Transient Facility telescope, surveys and data formats.
In Section \ref{sec:software} we outline the software stack we have developed to ingest and process lightcurve data for microlensing detection.
In Section \ref{sec:method} we step through the detection pipeline constructed to search for microlensing in ZTF's all sky survey data.
In Section \ref{sec:results} we analyze the results of our pipeline and share our list of microlensing candidates.
We conclude with a discussion in Section \ref{sec:discussion}.
\edit{Note, Sections \ref{sec:catterms} and \ref{sec:software} are not required to understand the methodology, results, and discussion presented in Sections \ref{sec:method}~-~\ref{sec:discussion}.}

\section{The Zwicky Transient Facility} \label{sec:telescope}

\subsection{Surveys and Data Releases} \label{sec:instrument}

The Zwicky Transient Facility (ZTF) began observing in March 2018 as an optical time-domain survey on the 48-inch Samuel Oschin Telescope at Palomar Observatory \citep{Bellm2019a, Graham2019}.
In the almost three years of Phase-I operations, ZTF has produced one of the largest astrophysical catalogs in the world.
Nightly surveys were carried out on a 47 square degree camera in ZTF g-band, r-band and i-band filters averaging $\sim2.0''$ FWHM on a plate scale of $1.01''\ \text{pixel}^{-1}$.
The surveys during this time were either public observations funded by the National Science Foundation's Mid-Scale Innovations Program, collaboration observations taken for partnership members and held in a proprietary period before later being released to the public, or programs granted by the Caltech Time Allocation Committee \citep{Bellm2019b}.

Three surveys of particular interest for microlensing science are the Northern Sky Survey and the Galactic Plane Survey, both public, and the partnership High-Cadence Plane Survey.
The Northern Sky Survey observed all sky north of $-31^{\circ}$ declination in g-band and r-band with an inter-night cadence of three days, covering an average of 4325 deg$^2$ per night.
The Galactic Plane Survey \citep{Prince_2018} observed all ZTF fields falling within the Galactic plane ($-7^{\circ} < b < 7^{\circ}$) in g-band and r-band every night that the fields were visible, covering an average of 1475 deg$^2$ per night.
These two public surveys have been run continuously since March 2018.
The collaboration High-Cadence Plane Survey covered 95 deg$^2$ of Galactic plane fields per night with 2.5 hour continuous observations in r-band totaling approximately 2100 deg$^2$.
ZTF Phase-II began in December 2020 with a public two-night cadence survey of g-band and r-band observations of the Northern sky dedicated to 50\% of available observing time.

The resulting data from these surveys is reduced and served to the public by the Infrared Processing and Analysis Center (IPAC) in different channels tailored to different science cases \citep{Masci2018}.
IPAC produces seasonal data releases (DRs) containing the results of the ZTF processing pipeline taken under the public observing time and a limited amount of partnership data.
The products included in these DRs include instrumentally calibrated single-exposure science images, both point-spread-function and aperture source-catalogs from these individual exposures, reference images constructed from a high quality set of exposures at each point in the visible sky, an objects table generated from creating source-catalogs on these reference images, and lightcurves containing epoch photometry for all sources detected in the ZTF footprint.
There have been DRs released every three to six months, starting with DR1 on 2019 May 8.
As discussed in more detail in Section \ref{sec:puzle}, this work exclusively uses data publicly available in DR3, DR4 and DR5 released on 2021 March 31.\footnote{
Additional details about the content and structures of products in these data releases can be found at \url{https://www.ztf.caltech.edu/ztf-public-releases.html}.}

These observations cover a wide range of Galactic structure in multiple filters with almost daily coverage over multiple seasons.
ZTF's capability to execute a wide-fast-deep-cadence opens an opportunity to observe microlensing events throughout the Galactic plane that microlensing surveys only focused in the Galactic bulge cannot observe with equivalent temporal coverage.
Our previous work simulating microlensing events observable by ZTF \citep{Medford2020b} indicated that there exists a large population of microlensing events both within and outside of the Galactic bulge yet to be discovered.
We seek to find black hole microlensing candidates amongst these events that could be later confirmed using astrometric follow up.

\subsection{Object Lightcurves}
\label{sec:object lightcurves}
While the real-time alert st\edit{r}eam is optimized for events with timescales of days to weeks, black hole microlensing events occur over months or even years.
Fitting for microlensing events requires data from both a photometric baseline outside of the transient event and the period during which magnification occurs ($t_0 \pm 2t_E$).
The data product therefore most relevant to our search for long-duration microlensing events are the data release lightcurves containing photometric observations for all visible objects within the ZTF footprint.

Data release lightcurves are seeded from the point spread function (PSF) source-catalogs measured from co-added reference images.
PSF photometry measurements from each single-epoch image are appended to these seeds where they occur in individual epoch catalogs.
A reference image can only be generated when at least 15 good-quality images are obtained, limiting the locations in the sky where lightcurves can be found to parts of the northern sky observed during good weather at sufficiently low airmass.
This sets the limit on the declination at which the Galactic plane is visible in DR lightcurves and consequently our opportunity to find microlensing events in these areas of the sky.
Statistics regarding the lightcurve coverage for nearly a billion objects have been collated from the DR5 overview into Table \ref{tab:dr4_stats}.

Observation fields are tiled over the night sky in a primary grid, with a secondary grid slightly shifted to cover the chip gaps created by the primary grid.
The lightcurves are written into separate lightcurve files, one for each field of the ZTF primary and secondary grids and spanning approximately $7^{\circ}\ \text{x}\ 7{^\circ}$ each.
In each one of these lightcurve files is a large ASCII table with a single row detailing each lightcurve metadata (including right ascension, declination, number of epochs, and so forth), followed by a series of rows containing the time (in heliocentric modified julian date, or hmjd), magnitude, magnitude error, linear color coefficient term from photometric calibration, and photometric quality flags for each single epoch measurement.
These lightcurve files are served for bulk download from the IPAC web server and total approximately 8.7 terabytes.

\edit{A detailed summary of terminology for the ZTF catalog follows in Section \ref{sec:catterms}. 
Here we provide a concise summary; see also Figure \ref{fig:puzle_diagram}.
We emphasize that lightcurve files and lightcurves are distinct. ``Stars" each have a unique RA and Dec and are assumed to be individual astrophysical objects. These ``stars" may be several astrophysical stars blended together. Each ``star" is associated with multiple ``sources" that have different fields and readout channels due to ZTF's observing patterns. These are stored in lightcurve files, which are ASCII files with many lightcurves that all share the same field. If a star is associated with multiple sources in the same field but with different readout channels, there will be multiple sources associated with the same star in one lightcurve file. Each source is associated with 1-3 ``objects" with each object having a unique filter; we call all the objects associated with one source ``siblings." Finally, each object is associated with a unique lightcurve.
}

\begin{table*}[t]
  \centering
    \begin{tabular}{c || cccc}
      \hline
      \hline
      Filter & Accessible Sky-Coverage & $N_\text{lightcurves}$ & $N_\text{lightcurves}$ with $N_{obs} \ge 20$ \\
      \hline
      g-band & 97.66\% & 1,226,245,416 & 582,677,216 \\
      r-band & 98.31\% & 1,987,065,715 & 1,107,250,253 \\
      i-band & 51.88\% & 346,398,848 & 78,425,164 \\
    \end{tabular}
    \caption{Object Statistics for Public Data Release 5. Additional details can be found at \href{https://www.ztf.caltech.edu/page/dr5}{https://www.ztf.caltech.edu/page/dr5}.}
    \label{tab:dr4_stats}
\end{table*}

\subsubsection{Catalog Terminology}
\label{sec:catterms}
\begin{figure*}[!htb]
    \centering
    \includegraphics[width=\textwidth, angle=0]{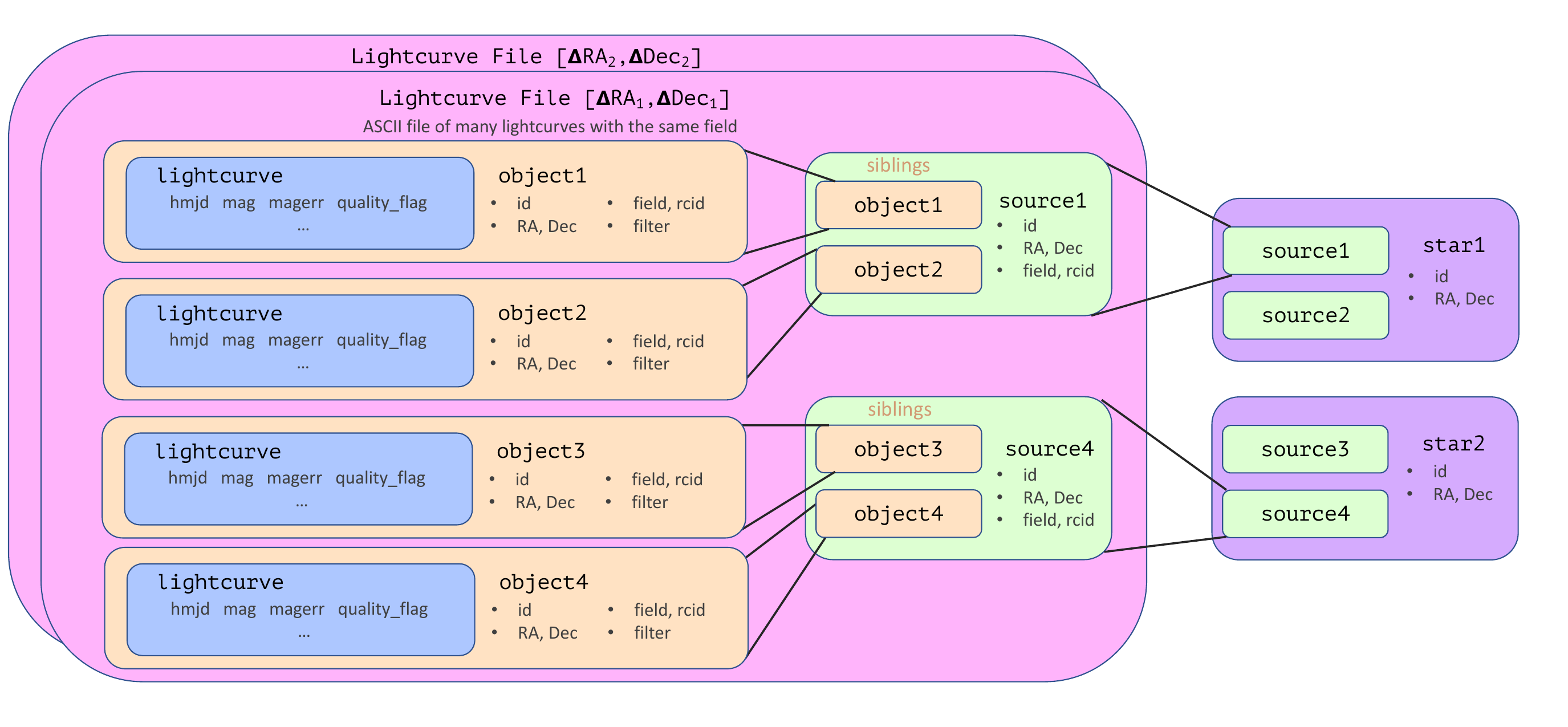}
    \caption{Diagram of how observational information is stored in the \texttt{PUZLE} pipeline. An object is a set of observations at the same sky position taken on the same field, readout channel and filter. A source can contain multiple objects from the same field and readout channel that are each different filters. \edit{We refer to multiple objects associated with the same source as siblings.} This distinction exists because observations in different filters are assigned different identification numbers in the ZTF public data releases. \edit{One star can have multiple sources across lightcurve files, so} all sources are cross-referenced with each other by sky location and together form a single star. \edit{A lightcurve file is an ASCII file containing all the lightcurves associated with the same field, so it contains lightcurves from many stars. If a star has multiple sources with the same field, but different readout channels due a shift in position caused by pointing errors, then the star will have multiple sources in the same lightcurve file.} \label{fig:puzle_diagram}}
\end{figure*}

Continuing to discuss the structure of the lightcurve files 
requires a consistent vocabulary for describing this data product \edit{which} we introduce here for clarity.
A diagram of the terms introduced here is shown in Figure \ref{fig:puzle_diagram}.
An object refers to a collection of photometric measurements of a single star in a single filter, in either the primary or secondary grid (but not both).
Objects are written in the lightcurve files\edit{, which are associated with a unique field,}
grouped by their readout channel \edit{into sources}.
A source refers to the total set of objects within a single lightcurve file 
that are all taken at the same sky location and are assumed to arise from the same astrophysical origin.
There can be a maximum of three objects per source for the three ZTF filters.
Crowding in dense galactic fields will often result in different astrophysical signals falling in the same PSF and therefore the same object. We address this issue during our microlensing fitting. 
Each of the objects within a single source are \textit{siblings} to one another and have mutually exclusive filters.
Siblings must share the same readout channel within a lightcurve file.
A star refers to all sources of the same astrophysical origin found in all of the lightcurve files, so the star will have multiple sources if it has been captured by both the primary and secondary grids or if the star appeared on multiple readout channels.
While many of these measurements will not \edit{be} stars but in fact several stars blended together, galaxies, or other astrophysical phenomena, we adopt this nomenclature to serve our purpose of searching for microlensing events mainly in the galactic plane.
Each star contains one or more sources, with each source containing one or more objects, with each object containing one lightcurve.

There are several limitations \edit{inherent} to the construction of these lightcurves that we have addressed in our work.
First, the lightcurves for each filter are created from independent extraction of each filter's single epoch flux measurement at a given location of the sky.
Flux measurements in the g-band, r-band, and i-band filters from a single astrophysical origin will have three separate object IDs at three different locations in the lightcurve file with no association information present relating the three objects to each other.
In terms of our nomenclature, it is unknown whether a source will have one, two, or three objects until the entire lightcurve file is searched row by row.

Second, objects found in different fields in the ZTF observing grid are also not associated with each other despite coming from the same astrophysical origin.
The ZTF secondary grid intentionally overlaps with the primary grid to increase coverage in the gaps between readout channels and fields found in the primary grid.
However this results in single filter measurements from the same star being split into two primary and secondary grid lightcurve files.
We have found instances where telescope pointing variations \edit{have} resulted in the same star having sources located in two different primary grid lightcurve files \edit{or the same star being recorded in multiple readout channels leading to multiple sources in one lightcurve file}.
Therefore it is unknown whether a single star will contain one or more sources without searching through both the primary and secondary grid lightcurve files, or searching adjacent lightcurve files for those stars that fall on the edges of each field.

Third, the format of the lightcurve files are raw ASCII tables.
While this format is exchangeable across multiple platforms and therefore reasonably serves the purpose of a public data release, ASCII is suited to neither high-speed streaming of the data from disk to memory nor searching for objects with particular properties without reading through an entire file.
If we are to search for microlensing events throughout the entire ZTF catalog then we must either ingest these lightcurves files into a more traditional database structure or create a method for efficiently reading these lightcurves files that provides the benefits of a database structure.

\section{Software} \label{sec:software}

We have developed a set of software tools to address the outstanding issues with the DR lightcurve data product.
These tools are a mixture of public open source codes and internal codes that enable efficient access to the data contained in the lightcurve files despite their ASCII format.
The functionality and availability of these tools proved essential to scaling out microlensing search method\edit{s} to the entire ZTF lightcurve catalog.
However, these tools and approaches would prove generally useful to any science that seeks to execute a large search across the entire set of bulk lightcurves.
We will therefore outline our approach in a level of technical detail that would aide another researcher attempting to similarly carry out such a large-scale search using the publicly available ZTF lightcurves.

\subsection{\texttt{zort}: ZTF Object Reader Tool} \label{sec:zort}

A reasonable initial approach to reading the lightcurve files would be to ingest all of the data into a relational database where objects could be cross-referenced against each other to construct sources and sources cross-referenced against each other to construct stars.
Flux measurements could be organized by object ID and an association table created to match these measurements to the metadata for each object.
However, this approach has two major drawbacks.
For science cases where the lightcurve files are used not as a search catalog but instead \edit{as} an historical reference for a singular object of interest, it is overkill to ingest all data into an almost 8 terabyte database.
Additionally, our approach aim\edit{s} to leverage the computational resources available at the NERSC Supercomputer located at the Lawrence Berkeley National Laboratory.
Here we can simultaneously search and analyze data from different locations in the sky by splitting our work into parallel processes.
We would experience either race conditions or the need to exclusively lock the table for each process' write if all of our processes were attempting to read and write to a singular database table.
Relational database access is a shared resource at NERSC, compounding this problem by putting a strict upper limit on the number of simultaneous connections allotted to single user\edit{s} when connecting to a database on the compute platform.
These bottlenecks would negate the benefits of executing our search on a massive parallel supercomputer.
We therefore sought a different method for accessing our data that would keep the data on disk and allow for so-called embarrassingly parallel data access.

The ZTF Object Reader Tool, or \texttt{zort}, is an open source Python package that serves as an access platform into the lightcurve files that avoids these bottlenecks \citep{zort}.
It is a central organizing principle of \texttt{zort} that keeping track of the file position of an object makes it efficient to locate the metadata and lightcurve of that object.
To enact this principle, \texttt{zort} requires four additional utility products for each lightcurve file.
These products are generated once by the user to initialize their copy of a data release by running the \texttt{zort-initialize} executable in serial or parallel using the \edit{P}ython \texttt{mpi4py} package.

During initialization, each line of a lightcurve file (extension \texttt{.txt}) containing the metadata of an object is extracted and placed into an objects file (extension \texttt{.objects}), with the file position of the object within the lightcurve file appended as an additional piece of metadata.
This file serves as an efficient way to blindly loop through all of the objects of a lightcurve file.
For each object in the objects file, \texttt{zort} will jump to the object's saved file position and load all lightcurve data into memory when requested.
Additionally, a hash table with the key-value pair of each object's object ID and file position is saved to disk as an objects map (extension \texttt{.objects\_map}).
This enables near-immediate access to any object's metadata and lightcurve simply by providing \texttt{zort} with an object ID.

Next, a k-d tree is constructed from the sky positions for all of the objects in each readout channel and filter within a lightcurve file and consolidated into a single file (extension \texttt{.radec\_map}).
This k-d tree enables the ability to quickly locate an object with only sky position.
Lastly, a record of the initial and final file position of each filter and readout channel within a lightcurve file is saved (extension \texttt{.rcid\_map}).
Lightcurve files are organized by continuous regions of objects that share a common filter and readout channel, and by saving this readout channel map we are able to limit searches for objects to a limited set of readout channels if needed.

\texttt{zort} presents the user a set of Python classes which enable additional useful features.
Lightcurve files are opened with the \texttt{LightcurveFile} class that supports opening in a \texttt{with} context executing an iterator construct for efficiently looping over an objects file by only loading objects into memory as needed.
This \texttt{with} context manager supports parallelized access with exclusive sets of objects sent to each process rank, as well as limiting loops to only specific readout channels using the readout channel map.
Objects loaded with the \texttt{Object} class contain an instance of the \texttt{Lightcurve} class.
This \texttt{Lightcurve} class applies quality cuts to individual epoch measurements as well as color correction coefficients if supplied with an object's PanSTARRS g-minus-r color.
Objects contain a \texttt{plot\_lightcurves} method for plotting lightcurve data for an object alongside the lightcurves of its siblings.
Each object has a \texttt{locate\_siblings} method that uses the lightcurve file's k-d tree to locate coincident objects of a different filter contained within the same field.
The package has a \texttt{Source} class for keeping together all sibling objects of the same astrophysical origin from a lightcurve file.
Sources can be instantiated with either a list of object IDs or by using the utility products to locate all of the objects located at a common sky position.
\texttt{zort} solves all problems related to organizing and searching for objects across the right ascension polar transition from $360^\circ$ to $0^\circ$ by projecting instrumental CCD and readout channel physical boundaries into spherical observation space and transforming coordinates.

\texttt{zort} has been adopted by the ZTF collaboration and TESS-ZTF project as a featured tool for extracting and parsing lightcurve files.
\texttt{zort} is available for public download as a GitHub repository (\href{https://github.com/michaelmedford/zort}{https://github.com/michaelmedford/zort}) and as a pip-installable PyPi package (\href{https://pypi.org/project/zort}{https://pypi.org/project/zort/}).

\subsection{\texttt{PUZLE}: Pipeline Utility for ZTF Lensing Events}
\label{sec:puzle}

Our search for microlensing events sought to combine all flux information from a single astrophysical origin by consolidating all sources into a single star.
This required a massive computational effort as flux information was scattered across different lightcurve files as different objects with independent object IDs.
First we identified all sources within each lightcurve file by finding all of the siblings for each object within that file.
Then sources in different lightcurve files at spatial coincident parts of the sky were cross-referenced with each other to consolidate them into stars.
To execute this method, as well as apply microlensing search filters and visually examine the results of this pipeline via a web interface, we constructed the \edit{P}ython Pipeline Utility for ZTF Lensing Events or \texttt{PUZLE}.
Similar to the motivations for constructing \texttt{zort}, this package needed to take advantage of the benefits of massive supercomputer parallelization without hitting the bottlenecks of reading and writing to a single database table.

\begin{figure*}[!htb]
    \centering
    \includegraphics[width=0.9\textwidth, angle=0]{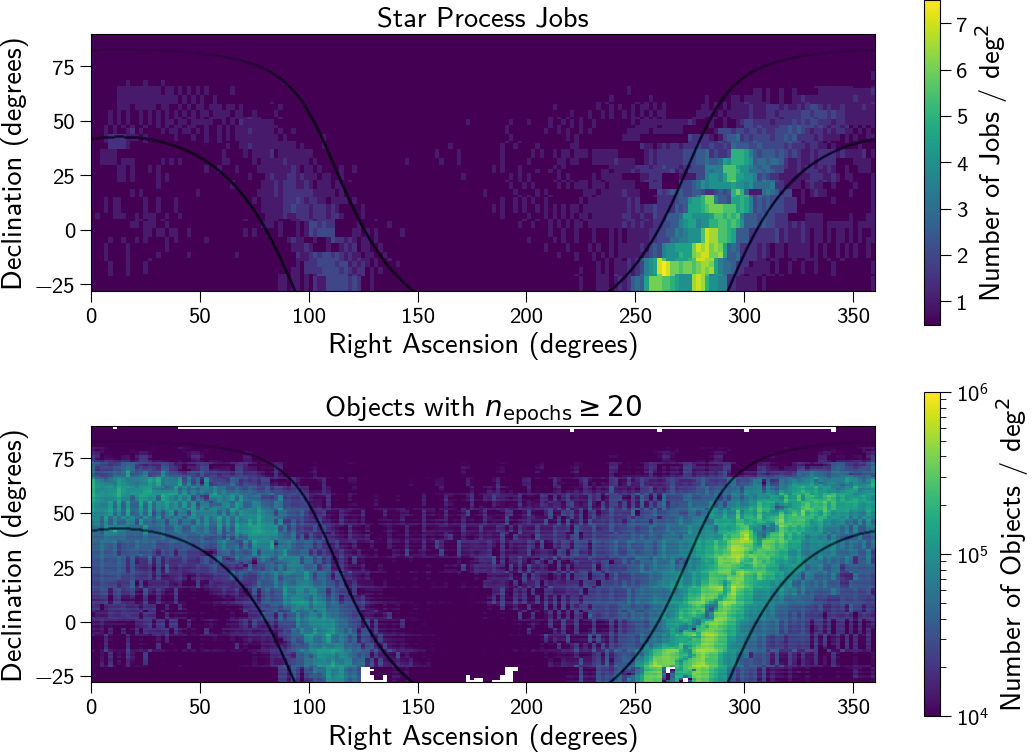}
    \caption{Processing jobs for the \texttt{PUZLE} pipeline (top) and density of objects with a minimum number of observations (\edit{with \textit{catflag} $<$ 32768}; bottom) throughout the sky, with the boundaries of the Galactic plane outlined in black. Job cells were sized to enclose approximately the same number of sources resulting in a larger number of smaller sized jobs in the areas of the sky with more objects per square degree. All job cells had a fixed height of $1^{\circ}$ and a variable width between $0.125^{\circ}$ and $2.0^{\circ}$. The limitations of a Galactic survey from the Northern hemisphere are also seen in the lack of sources near the Galactic center. \label{fig:cells}}
\end{figure*}

The \texttt{PUZLE} pipeline began by dividing the sky into an grid of adaptively sized cells that contain approximately equal number of stars.
Grid cells started at $\delta=-30^{\circ}, \alpha=0^{\circ}$ and were drawn with a fixed height of $\Delta \delta = 1^{\circ}$ and a variable width of $0.125^{\circ} \le \Delta \alpha \le 2.0^{\circ}$ at increasing values of $\alpha$.
We attempted to contain no more than 50,000 objects within each cell using a density determined by dividing the number of objects present in the nearest located lightcurve file by the file's total area. 
The cells were drawn up to $\alpha = 360^{\circ}$, incremented by $\Delta \delta = 1.0^{\circ}$, and repeated starting at $\alpha = 0^{\circ}$, until the entire sky had been filled with cells.
The resulting grid of 38,819 cells can be seen in Figure \ref{fig:cells}.
Each of these cells represents a mutually exclusive section of
the sky that was split between parallel processes for the construction of sources and stars.

A PostgreSQL database was created to manage the execution of these parallel processes.
This table was vastly smaller than a table that would contain all of the sources and stars within ZTF and was therefore not subject to the same limitations as previously stated for applying relational databases to our work.
We created two \texttt{identify} tables, each containing a row for each cell in our search grid, one for identifying sources and one for identifying stars.
Each row represented an independent \texttt{identify\_job} that was assigned to a compute core for processing. 
Each row contained the bounds of the cell, and two boolean columns for tracking whether a job had been started and/or finished by a compute core.
Historical information about the date and unique compute process ID was also stored in the row for debugging purposes.
To ensure that per-user database connection limits were not exceeded, we used an on-disk file lock.
A parallel process had to be granted permission to this file lock before it attempted to connect to the database, thereby offloading the bottleneck from the database to the disk.

Here we describe how each \texttt{identify\_job} script worked, whether it is identifying sources or stars.
An \texttt{identify\_job} script was submitted to the NERSC supercomputer requesting multiple compute cores for a fixed duration of time.
A script began by each compute core in the script fetching a job row from the appropriate \texttt{identify} table that was both un-started and un-finished and marked the row as started.
The script then identified all of the sources (or stars) within the job bounds as described below.
The resulting list of sources (or stars) was then written to disk for later processing.
Lastly the job row in the \texttt{identify} table was marked as finished.
Each script continued to fetch and run \texttt{identify\_job}s for all of its compute cores until just before the compute job was set to expire.
At this time it interrupted whichever jobs were currently running and reset their job row to mark them as un-started.
In this way the \texttt{identify} table was always ready for any new \texttt{identify\_job} script to be simultaneously run with any other script and ensure that only un-started and un-finished jobs were run.
The \texttt{identify\_job} script for stars could only request jobs where the associated script for sources had finished.

The \texttt{identify\_job} for finding sources began by finding the readout channels within all of the lightcurve files that intersected with the spatial bounds of its job by using the projected readout channel coordinates calculated by the \texttt{zort} package.
Special attention was paid here to lightcurve files that cross the right ascension polar boundary between $360^\circ$ and $0^\circ$ to ensure that they were correctly included in jobs near these boundaries.
Objects within the readout channels that overlap with the job were then looped over using the \texttt{zort LightcurveFile} class and those objects outside of the bounds of the job were skipped.
Objects with fewer than 20 epochs of good quality observations were also skipped over as this cut is an effective way to remove spurious lightcurves arising from erroneous or non-stationary seeds (see Figure 3 on the \href{https://irsa.ipac.caltech.edu/data/ZTF/docs/releases/dr05/ztf_release_notes_dr05.pdf}{DR5 page}).
The remaining objects had their siblings located using the \texttt{locate\_siblings} method and all of the siblings were grouped together into a single source.
Once all of the lightcurve files had been searched, the remaining list of sources was cut down to only include unique sources with a unique set of object IDs.
This prevented two duplicate sources from being counted, such as when a g-band object pairs with a r-band sibling, and that same r-band object pairs with that same g-band object as its sibling.
Each source was assigned a unique source ID.
The object IDs and file positions of all objects within each source were then written to disk as a source file named for that \texttt{identify\_job}.
In addition to this source file, a hash table similar to the \texttt{zort objects\_map} file was created.
This source map was a hash table with key-value pairs of the source ID and file position within the source file where that source was located.

The \texttt{identify\_job} for finding stars began by loading all of the sources for that job directly from the on-disk file written by the source script.
The sky coordinate of each source was loaded into a k-d tree.
Each source's neighbors located within 2 arcseconds were found by searching through this tree.
Sources were grouped together to form a star.
A star's location was calculated as the average sky coordinate of its sources and the star defined as the unique combination of source IDs which it contains.
Each star was assigned a unique ID.
The star id, sky location and list of source IDs of each star were written to disk in a star file.
A star map similar to the source map described above was also written to disk.

\subsection{Upgrading to a New Public Data Release}

Our approach of using the file position to keep track of objects and sources was able to transition between different versions of the public data releases when available.
We used an older DR as a seed release upon which we initially searched for sources and stars.
For this microlensing search and by way of example, we used DR3 as a seed release and updated our source and star files to DR5.
We began by creating the necessary \texttt{zort} utility products for DR5 lightcurve files.
We sought to append additional observations to our seed list of sources and stars.
However, we did not look to find any additional objects that did not exist in DR3.
The convention for different DRs is to keep the same object IDs for reference sources found at the same sky coordinates and to add new IDs for additional objects found.
Therefore when updating a row in the source file from DR3 to DR5 we only needed to use the source's object IDs and the DR5 object map to find the file location of the object in the DR5 lightcurve file.
We found that in this process less than 0.02\% of the object IDs in a source file could not be located in the DR5 object map and were dropped, resulting in a 99.98\% conversion rate that was acceptable for our purposes.
A new DR5 source map was then generated for this DR5 source file.
Star files did not need to be updated and could be simply copied from an older to a newer data release.
It should be noted that the names of lightcurve files slightly change between data releases, as the edge sky coordinates are printed into the file names and are a function of the outermost object found in each data release for that field.
Any record keeping that involved the names of these lightcurve files needed to be adjusted accordingly.

\section{Detection Pipeline} \label{sec:method}

With our pipeline, we sought to measure the $t_E$ distribution for long-duration microlensing events ($t_E \geq 30$ days) in order to search for a statistical excess due to black hole microlensing.
Second, we sought to generate a list of black hole candidates that could be followed up astrometrically in future studies.
We therefore made design choices optimized for increasing the probability of detecting these types of events, even at the expense of short-duration events. 
We designed our pipeline to remove the many false positives that a large survey like ZTF can generate, even at the expense of false negatives.
Our detection pipeline had to be capable of searching through an extremely large number of objects, sources and stars and therefore had to be both efficient and computationally inexpensive to operate.
These two priorities informed each of our decisions when constructing our microlensing detection pipeline. \edit{In our pipeline, we make progressively stricter cuts which yields ``levels" of candidates between 0 (no cuts) and level 6 (strictest cuts). See Table \ref{tab:pipeline_cuts} for a summary of cuts made at each level and the number of remaining candidates.}

\subsection{\texttt{process} Table}
In order to efficiently distribute and manage parallel analysis processes on a multi-node compute cluster like NERSC, we utilized a \texttt{process} table. 
This table contains a row for each cell in our search grid, just as we did for the sources and stars.
However, each row also had columns for detection statistics that we kept track of throughout the pipeline's execution.
\edit{Metadata was also saved to record thresholds that the algorithms automatically determined, which was useful for later debugging and analysis.}
We reiterate that the cost of reading and writing to such a table from many parallel processes is not a constraining factor when each job needs to only read and write to this table once.

The pipeline continued with a \texttt{process\_job} script selecting a job row from the \texttt{process} table that was un-started and un-finished.
The job read in the star file for its search cell and the associated source maps for each lightcurve file from which a source could originate within its cell.
These source maps were used to find the source file locations for each source ID of a given star.
Each source row in the source file had the object IDs of the source's objects and \texttt{zort} could use these object IDs to load all object and lightcurve data into memory.
Throughout the pipeline we maintained a list of stars and matched that list with the list of sources associated with each of those stars.
Our pipeline performed calculations and made cuts on objects, but our final visual inspection was done on the stars to which those objects belong.
We therefore needed to keep track of all associations between stars, sources, and objects throughout the pipeline.
All cuts described below were performed within the stars, sources and objects of each \texttt{process\_job}.

\subsection{Level 1: Cutting on Number of Observations and Nights}
\label{sec: level 1}
The first cut we implemented was to \edit{remove all objects with fewer than 20 epochs of good data to avoid spurious lightcurves (see Section \ref{sec:puzle} for details).} \edit{This cut was made by cutting on a \textit{catflag} of $<$ 32768 as described in Section 13.6 of \href{https://irsa.ipac.caltech.edu/data/ZTF/docs/ztf_explanatory_supplement.pdf}{The ZTF Science Data System (ZSDS)
Explanatory Supplement}. The \textit{catflag} describes the data quality and if the measurement has a specific issue; if the $catflag \geq 32768$, it contains bit 15 which means it is likely affected by the moon or clouds, so if it does not contain that bit, the data is ``probably usable". Filtering just on bit 15 does not lead to ``perfectly clean" extractions, which requires \textit{catflag = 0}; we made that cut later (see Section \ref{sec: level 4}).} \edit{We also} removed any objects with less than 50 unique nights of observation.
There were fields in our sample that were observed by the High-Cadence Plane Survey that contain many observation epochs but all within a few nights \citep{Bellm2019b}.
We limited our search to those events with many nights of observation due to the long duration of black hole microlensing events, as described in Section \ref{sec:intro}.
While we had already performed a cut on objects with less than 20 good quality observations, performing this cut removed those objects which only passed our initial cut due to a small number of nights being sampled multiple times.
We were left with 1,011,267,730 objects and 563,588,562 stars in our level 1 catalog.

\subsection{Level 2: Cutting on the von Neumann ratio, Star Catalogs, and a Four-Parameter Microlensing Model}
\label{sec: level 2}
\subsubsection{Cutting on the von Neumann ratio}
\citet{PriceWhelan2014} developed a detection method for finding microlensing events in large, non-\edit{uniformly} sampled time domain surveys using Palomar Transient Factory (PTF) data.
Their key insight was to use $\eta$, or the von Neumann ratio (also referred to as the Durbin-Watson statistic; \citealt{vonNeumann1941, Durbin1971}), to identify microlensing events.
This statistic is an inexpensive alternative to the costly $\Delta \chi^2$ that measures the difference in $\chi^2$ of fitting data to a flat model and to a microlensing model.
Their pipeline was also biased toward removing false positives at the expense of false negatives by culling their data on statistical false positive rate thresholds for $\eta$.
ZTF captures nearly an order of magnitude more sky coverage with each exposure than PTF with different systematics, motivating a different implementation of this statistic.

\begin{figure*}[!htb]
    \centering
    \includegraphics[width=0.9\textwidth, angle=0]{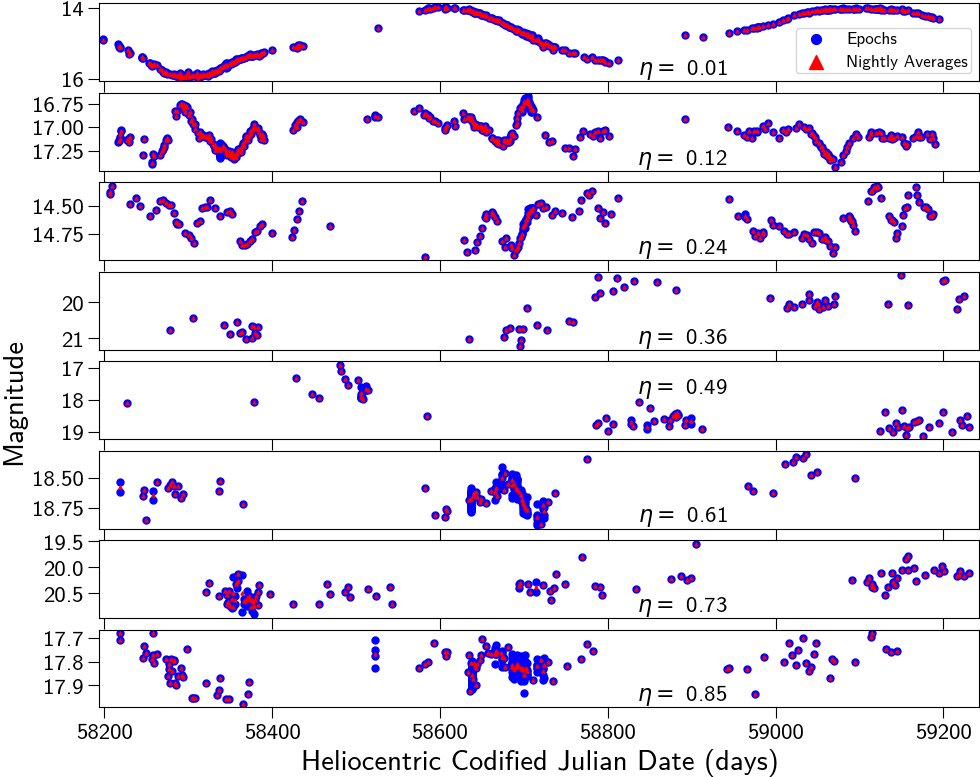}
    \caption{Example ZTF objects and their $\eta$ values, with individual epochs (blue) and nightly averages (red). DR5 lightcurves contained various cadences and gaps in the data depending on their visibility throughout the year and \edit{on} which ZTF surveys were executed at their location. This resulted in a heterogeneous dataset that required a flexible approach. The $\eta$ statistic was able to capture inter-epoch correlation despite these different observing conditions, with smaller $\eta$ signaling more correlated variability in the lightcurve. \edit{Note that the data plotted was cleaned using \textit{catflag} $<$ 32768.}  \label{fig:example_lightcurves}}
\end{figure*}

We calculated the von Neumann ratio $\eta$ on all objects in the level 1 catalog:
\begin{align}
    \eta = \frac{\delta^2}{\sigma^2} = \frac{\Sigma_i^{N-1} (x_{i+1} - x_i)^2} {N - 1}\frac{1}{\sigma^2}.
\end{align}
$\eta$ is the ratio of the average mean square difference between a data point $x$ and its successor $x$+1 to the variance of that dataset $\sigma$.
Highly correlated data has a small difference between successive points relative to the global variance and have a correspondingly small $\eta$.
Gaussian noise has an average $\eta \approx 2$ with a smaller variance in the measurement for datasets with more points.
Several example lightcurves and their associated $\eta$ values are demonstrated in Figure \ref{fig:example_lightcurves}.

We calculated our $\eta$ not on an object's epoch magnitudes, but instead on an object's nightly magnitude averages.
Fields observed by the High-Cadence Plane Survey were observed every 30 seconds and had additional correlated signal due to being sampled on such short time scales relative to the dynamic timescale of varying stellar brightness.
This biased the objects in a job cell toward lower $\eta$.
Calculating $\eta$ on nightly magnitude averages removed this bias and created an $\eta$ distribution closer to that expected by Gaussian noise.
This reduced the undue prominence of those objects observed by the High-Cadence Plane Survey when searching for long-duration microlensing events.
We performed all $\eta$ calculations on the dates and brightness of nightly averages.

A cut on $\eta$ required a false positive threshold that separated those events with significant amounts of correlated signals and those without.
\edit{Our calculation of this threshold was based on \citet{PriceWhelan2014} with two alterations.}
First we chose to calculate a threshold not from determining the 1\% false positive recovery rate for scrambled lightcurves, but instead by finding the $1^{\text{st}}$ percentile on the distribution of $\eta$.
Initial attempts to set the threshold from scrambled lightcurves resulted in more than 1\% of the objects passing our cut due to correlated noise not explained by the global variance.
Only passing the 1\% of objects with the lowest $\eta$ guaranteed that this stage of the pipeline would remove 99\% of events, significantly cutting down on the number of objects passing this stage of the pipeline.

Second, we chose to bin our data by number of observation nights and calculated a separate threshold for each of these bins.
The variance of $\eta$ is correlated with the number of data points in its measurement and therefore so too was the $1^{\text{st}}$ percentile of $\eta$ correlated with the number of observations.
Objects in our level 1 catalog span from 50 nights of observation to nearly 750 nights due largely to the presence of both primary and secondary grid lightcurves, as well as the mixture of ZTF public and partnership surveys, falling into the same job cell.
A single threshold calculated from all lightcurves would be biased toward passing short duration lightcurves and removing long duration lightcurves.
Our binning significantly dampened the effect of this bias by only comparing lightcurves with similar numbers of observation nights to each other.
In order to efficiently divide the lightcurves into these bins, we determined the cumulative distribution function for number of observation nights 
and defined the bin edges at the points where cumulative distribution function equaled 0.33 and 0.66.
These bin edges were unique for each job and were recorded in the \texttt{process} table.
Each lightcurve was then compared to these bin edges and assigned to the appropriate bin.
For each bin we calculated the $1^{\text{st}}$ percentile of $\eta$ and removed all lightcurves with $\eta$ greater than this threshold.
We also calculated the $90^{\text{th}}$ percentile of $\eta$ in each bin and saved it for a later stage in our pipeline.
\subsubsection{Cutting on the PS1-PSC Star Catalog}
Our next cut was a star-galaxy cut on those sources which we were confident are not astrophysical stars (not to be confused with our nomenclature for the word ``star'').
We used the Probabilistic Classifications of Unresolved Point Sources in PanSTARRS1 (PS1-PSC) that classified $\sim$1.5 billion PanSTARRS1 sources as either extended sources or point sources using a machine learning model \citep{Tachibana2018}.
Each remaining object in a job was queried against this catalog to find a corresponding PS1-PSC score at that location in the sky.
Lightcurves were retained that had a PS1-PSC score greater than or equal to 0.645, or did not have a corresponding score.
This threshold is the value at which PS1-PSC labeled sources with a \texttt{rKronMag}$< 21$ (which captures nearly all ZTF sources) as astrophysical stars with a 96.4\% true positive rate and only costs a false positive rate of 1.0\%.
This cut retained 96.4\% of the astrophysical stars in our sample while only permitting 1.0\% of galaxies to pass.
To avoid the same database bottlenecks previously described, we downloaded the entire PS1-PSC catalog onto disk (private communication Adam Miller) broken into individual files for separate sections of the sky.
We generated k-d trees for each of these files and all spatially coincident PS1-PSC catalog files were loaded into memory at the beginning of each \texttt{process\_job}, enabling a fast PS1-PSC score look up for each object at run-time.
\subsubsection{Cutting on a Four-Parameter Microlensing Model}
Lastly we fit a four-dimensional microlensing model to the daily average magnitudes of each remaining object.
Microlensing models are multi-dimensional and non-linear, which results in costly fitting that would be prohibitive to a search of this scale.
\citet{Kim2018} outlined an analytical representation of microlensing events that circumvented this issue by only attempting to fit microlensing events in the high magnification ($u_0 \lesssim 0.5$) and low magnification ($u_0$=1) limits.
The microlensing model representation they deduced for these limits was
\begin{equation}
\begin{aligned}
    F(t) &= f_1 A_j (Q(t; t_0, t_\text{eff})) + f_0; \\
\end{aligned}
\end{equation}
\edit{to model the flux over time with four free parameters: $f_0, f_1, t_0,\ \text{and}\ t_\text{eff}$. The amplification from microlensing is $A_j$ which is a function of $Q$, with $j=1,2$ corresponding to $u_0 \lesssim 0.5$ and $u_0$=1 limits, respectively,
and is defined as}
\begin{eqnarray}
    A_{j=1}(Q) &=& Q^{-1/2} \\
    A_{j=2}(Q) &=& [1 - (Q/2 + 1)^{-2}]^{-1/2} \\
    Q(t; t_0, t_\text{eff}) &\equiv& 1 + \left( \frac{t - t_0}{t_\text{eff}} \right)^2.
\end{eqnarray}
In this construction $f_0$ and $f_1$ no longer have a physical interpretation \edit{\citep[as defined in ][]{Kim2018}} but are instead simply parameters of the fit.
$t_\text{eff}$ is a non-physical substitution for $t_E$ and $t_0$ is the time of the photometric peak.
At this stage in our pipeline we had sufficiently small numbers of events as to permit the simultaneous four dimensional fit of $f_0, f_1, t_0,\ \text{and}\ t_\text{eff}$ for both the low and high magnification limits.
Unlike the grid search that \citet{Kim2018} performed for a solution, our two fits were performed on each object with bounds (Table \ref{tab:fit_bounds}) and least squares minimization with the Trust Region Reflective algorithm \citep{Branch1999}.
The average time to fit each object to both the low and high magnification solution was $60 \pm 5$ milliseconds.
For each of these two fits the $\Delta \chi^2$ between the microlensing model and a flat model was calculated and the solution with the largest $\Delta \chi^2$ was kept as the best fit.
This model was subtracted from the lightcurve and $\eta_\text{residual}$ was calculated on these residuals.
If $\eta_\text{res}$ was low then the microlensing model had failed to capture the variable signal that allowed the lightcurve to pass the first $\eta$ cut and additional non-microlensing variability remained.
We only retained those lightcurves with little remaining variability in the residuals by removing all objects with $\eta_\text{residual}$ less than the $90^{\text{th}}$ percentile threshold calculated earlier in our first $\eta$ cut.

\begin{table}[t]
  \centering
  \caption{Boundaries for four-parameter microlensing model.}
    \begin{tabular}{c|cccc}
       & $t_0$ (hmjd) & $t_\text{eff}$ (days) & $f_0$ (flux) & $f_1$ (flux)\\
      \hline
      Low Bound & min($t$) - 50 & 0.01 & $-\infty$ & 0 \\
      High Bound & max($t$) + 50 & 5000 & $\infty$ & $\infty$ \\
    \end{tabular}
    \label{tab:fit_bounds}
\end{table}

At this stage we combined our different epoch bins to obtain a single list of objects that had passed all of our cuts, as well as the record of the sources and stars to which these objects belong.
This smaller catalog could be saved into a more conventional relational database as it no longer required massive parallel compute power to perform further cutting and fitting.
For each star, hereafter referred to as a candidate, all of the candidate information, including the list of all its source IDs, was uploaded to a \texttt{candidates} table.
If there were multiple objects belonging to the candidate that passed all cuts, the pipeline data of the object with the most number of nights was saved in the candidate database row.
In this way each candidate contains the information of the star from which it was derived but has a single object upon which further cuts can be performed.
The job row in the \texttt{process} table was marked as finished and updated with metadata relating to the execution of the job. 
These cuts reduced the number of objects from 563,588,562 to 7,457,583 level 2 candidates, as outlined in Table \ref{tab:pipeline_cuts}. 

\begin{table*}[t]
  \centering
    \begin{tabular}{cllrr}
      & \textbf{Cut Performed} & & \textbf{Objects Remaining} & \textbf{Stars Remaining} \\
      \hline
      \multicolumn{5}{|l|}{\textbf{Level 0}} \\
     
      \hline \\
      & ZTF DR5 lightcurves with $N_\text{obs} \geq 20$ & & 1,768,352,633 & - \\
      & $N_\text{obs} \geq 20$ & & 1,744,425,342 & 702,431,964 \\
      & $N_\text{nights} \geq 50$ & & 1,011,267,730 & 563,588,562 \\
      \hline
      \multicolumn{5}{|l|}{\textbf{Level 1 \hfill 563,588,562}} \\
      \hline \\
      & $\eta \leq 1^\text{st}$ Percentile & & - & 10,227,820 \\
      & PS1-PSC $\geq 0.645$ & & - & 8,987,330 \\
      & Successful 4-parameter model fit & & - & 8,749,737 \\
      & Duplicates between fields and filters removed & & - & 7,457,583 \\
      \hline
      \multicolumn{5}{|l|}{\textbf{Level 2 \hfill 7,457,583}} \\
      \hline \\
      & $\eta_\text{residual}\ \geq \eta\ * 3.82 - 0.077$ & & - & 92,201 \\
      \hline
      \multicolumn{5}{|l|}{\textbf{Level 3 \hfill 92,201}} \\
      \hline \\
      & N$_\text{nights,~catflag = 0} \geq$ 50 & & - & 90,810 \\
      & Successful 7-parameter model fit & & - & 90,232 \\
      & $\chi^2_\text{red, opt} \leq 4.805$ & & - & 49,376 \\
      & $2t_E$ baseline outside of $t_0 \pm 2t_E$ & & - & 14,854 \\
      & $\chi^2_\text{red, flat} \leq 3.327$  & & - & 13,273 \\
      & $\pi_{E \text{, opt}} \leq 1.448$   & & - & 11,480 \\
      & $t_0 - t_E \geq 58194$ & & - & 8,649 \\
      \hline
      \multicolumn{5}{|l|}{\textbf{Level 4 \hfill 8,649}} \\
      \hline \\
        & Successful Bayesian model fit & & - & 8,646 \\
        & $\chi_{red}^2 \leq 3$ & & - & 6,100 \\
        & $t_0 - t_E \geq 58194$ & & - & 5,496 \\
      \hline
      \multicolumn{5}{|l|}{\textbf{Level 4.5 \hfill 5,496}} \\
      \hline
      & & $\swarrow$ &$\searrow$ \textcolor{white}{spacespacespacespace} & \cr
        & \textbf{Cut Performed} & \textbf{Stars Remaining} & \textbf{Cut Performed} & \textbf{Stars Remaining} \\
        \hline
        & $t_0 + t_E > 59243$ & \hfill 1,558 & $t_0 + t_E \leq 59243$ & 3,938 \\
        \cline{1-3}
        \multicolumn{3}{|l|}{\textbf{Level Ongoing \hfill 1,558}} & & \\
        \cline{1-3}
        & & & $\sigma_{t_E} / t_E \leq 0.20$ & 2,428 \\
        & & & $|u_0| \leq 1.0$ & 1,711 \\
        & & & $b_\text{sff} \leq 1.5$ & 1,711 \\
        & & & $4t_E$ baseline outside of $t_0 \pm 2t_E$ & 950 \\
      
      \cline{4-5}
        & & & \multicolumn{2}{|l|}{\textbf{Level 5 \hfill 950}} \\
       \cline{4-5} \\
      & & & Manually assigned clear microlensing label & 66 \\
      & & & Unmatched to known objects & 60 \\
      \cline{4-5}
        & & & \multicolumn{2}{|l|}{\textbf{Level 6: Events \hfill 60}} \\
       \cline{4-5}
    \end{tabular}
    \caption{Cuts for the \texttt{PUZLE} pipeline. Catalogs are defined as the collection of candidates remaining after previous cuts (i.e. 8649 candidates in the level 4 catalog). Cuts between level 0 and level 3 (\edit{Sections \ref{sec: level 1} - \ref{sec: level 3}}) are applied to all objects within a star. Cuts between level 3 and level 4 (\edit{Section \ref{sec: level 4}}) are applied on the object within each star with the most number of observations. Cuts between level 4 and level 6 (\edit{Sections \ref{sec: level 4.5} - \ref{sec: level 6}}) are applied to all objects within the star that are fit by the Bayesian model.}
    \label{tab:pipeline_cuts}
\end{table*}

\subsection{Level 3: Cutting on $\eta_{\rm residual}$}
\label{sec: level 3}
\subsubsection{Simulated Microlensing Events}

Reducing the number of candidates further required knowing how potential microlensing events within the sample would be affected by particular cuts. \edit{This sample of simulated microlensing events will be used in level 3 and 4.}
We therefore generated artificial microlensing events and injected them into ZTF data.
Following the method outlined in \citet{Medford2020b}, we ran 35 \texttt{PopSyCLE} simulations \citep{Lam2020} throughout the Galactic plane and imposed observational cuts mimicking the properties of the ZTF instrument.
The Einstein crossing times and Einstein parallaxes of artificial events were drawn from distributions that were fit for each \texttt{PopSyCLE} simulation.
The impact parameters were randomly drawn from a uniform distribution between $[-2, 2]$.
The source flux fraction, or the ratio of the flux originating from the un-lensed source to the total flux observed from the source, lens and neighbors, was randomly drawn from a uniform distribution between $[0, 1]$.
These parameters were then run through a point-source point-lens microlensing model with annual parallax and only sets of parameters with an analytically calculated maximum source amplification greater than 0.1 magnitudes were kept.

\begin{figure*}[!htb]
    \centering
    \includegraphics[width=0.9\textwidth, angle=0]{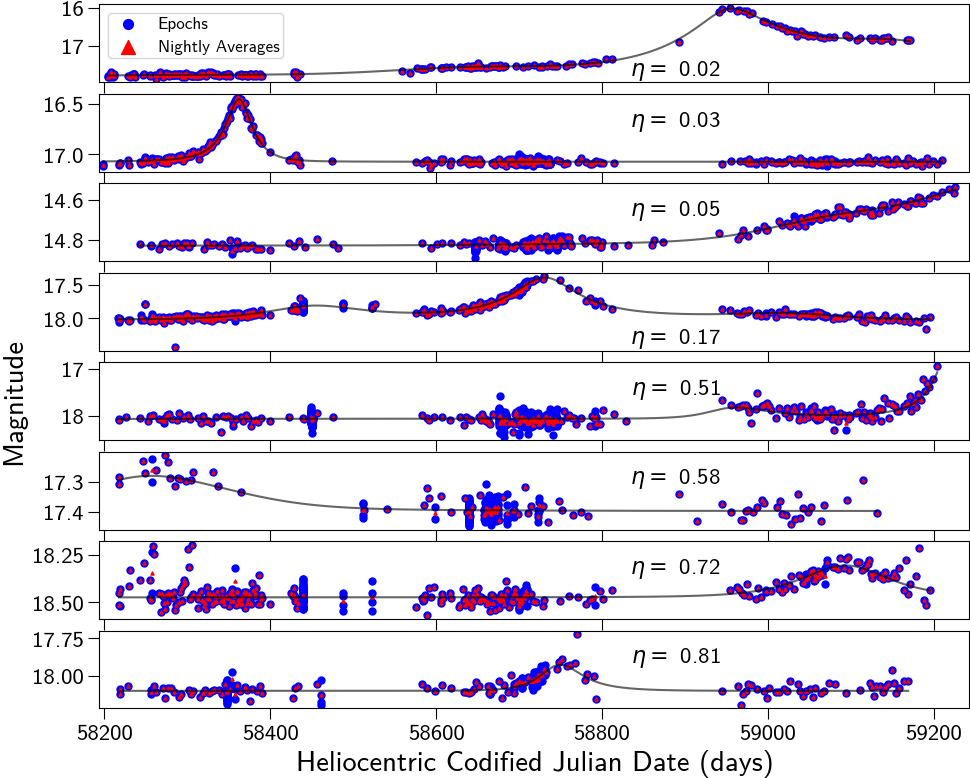}
    \caption{Example simulated microlensing lightcurves and their $\eta$ values, with individual epochs (blue) and nightly averages (red)\edit{, with data cut on \textit{catflag} $<$ 32768}. Each model (black line) was generated from distributions set by \texttt{PopSyCLE} simulations throughout the Galactic plane. The models were then injected directly into DR5 lightcurves, enabling calculations and cuts on these lightcurves to represent the diversity of cadence and coverage seen throughout the DR5 dataset. \label{fig:ulens_lightcurves}}
\end{figure*}

\edit{In order to generate simulated microlensing events with real noise, we injected our simulated events into random ZTF lightcurves.} We selected a sample of ZTF objects with at least 50 nights of observation located within the footprint of each corresponding \texttt{PopSyCLE} simulation.
This real lightcurve equals, in our model, the total flux from the lens ($F_\text{L}$), neighbors ($F_\text{N}$), and source ($F_\text{S}$) summing to a total $F_\text{LNS}$ outside of the microlensing event.
The light from the lens and neighbor can be written, using the definition of the source flux fraction $b_\text{sff} = F_\text{S} / F_\text{LSN}$, as
\begin{eqnarray}
	F_\text{LN} &=& F_\text{LSN} - F_\text{S}\ , \nonumber \\
	F_\text{LN} &= &F_\text{LSN} - b_\text{sff} \cdot F_\text{LNS}\ , \nonumber \\
	F_\text{LN} &=& (1 - b_\text{sff}) \cdot F_\text{LSN}\ .
	\label{eq:flux_ln}
\end{eqnarray}
An artificial microlensing lightcurve ($F_\text{micro}$) has amplification applied to only the flux of the source,
\begin{align}
	F_\text{micro} &= A \cdot F_\text{S} + F_\text{LN}\ .
	\label{eq:flux_micro_orig}
\end{align}
Equations \ref{eq:flux_ln} and \ref{eq:flux_micro_orig} can be combined to obtain a formula for the microlensing lightcurve using only the original ZTF lightcurve and the amplification \edit{and blending} of the simulated microlensing model,
\begin{align}
    	F_\text{micro} &= A \cdot b_\text{sff} \cdot F_\text{LSN} + (1 - b_\text{sff}) \cdot F_\text{LSN}\ .
	\label{eq:flux_micr}
\end{align}
A value of $t_0$ was randomly selected in the range between the first and last epoch of the lightcurve.
Events were then thrown out if they didn't have (1) at least three observations of total magnification greater than 0.1 magnitudes, (2) at least three nightly averaged magnitudes observed in increasing brightness in a row, and (3) have at least three of those nights be three-sigma brighter than the median brightness of the entire lightcurve.
This selection process yielded 63,602 simulated events with varying signal-to-noise as shown in Figure \ref{fig:ulens_lightcurves}.

\subsubsection{Computing $\eta_\text{residual}$ cut}
\begin{figure*}[!htb]
    \centering
    \includegraphics[width=0.9\textwidth, angle=0]{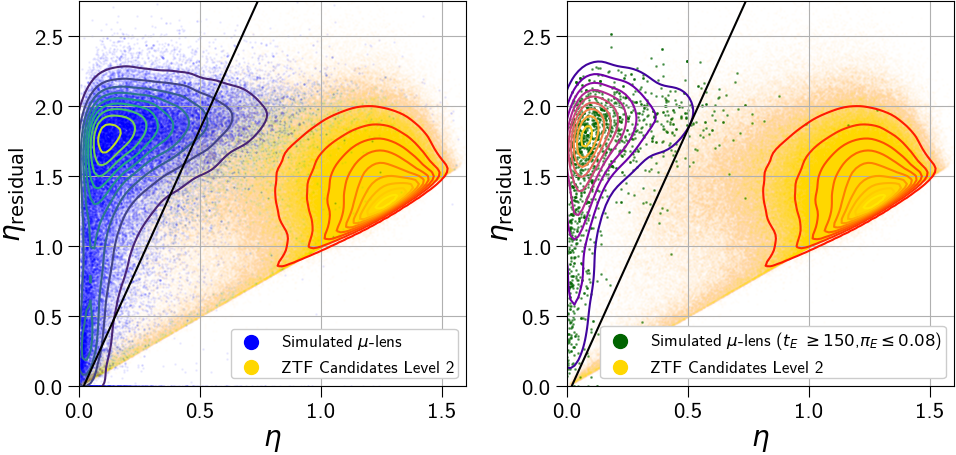}
    \caption{Microlensing events were clearly delineated in the level 2 catalog by calculating $\eta$ on lightcurves and the residuals of those lightcurves after fitting them \edit{to} a four-parameter microlensing model. \edit{Those data were cut on \textit{catflag} $<$ 32768.} $\eta$ was calculated on the nightly averages of ZTF candidates (yellow), simulated microlensing lightcurves (blue) and simulated microlensing lightcurves in the black-hole search space (green). This measurement of $\eta$ was made on both the entire lightcurve (x-axis) and on a lightcurve with the four-parameter model subtracted (y-axis). Microlensing events tended toward smaller $\eta$ values due to their correlated lightcurves and larger $\eta_\text{residual}$ when microlensing is the only source of variability. This enabled a cut (black line) in this $\eta$-$\eta_\text{residual}$ space that retained 83.6\% of simulated events and 95\% of black hole parameter space, while removing 98.8\% of the ZTF candidates from our sample. \label{fig:eta_eta_residual}}
\end{figure*}

$\eta$ and $\eta_\text{residual}$ were calculated on all simulated microlensing events.
Most of the simulated microlensing events had low values of $\eta$, due to the correlation between subsequent data points as compared to the sample variance, and high values of $\eta_\text{residual}$, due to the lack of correlation after subtracting a successful microlensing model.
Figure \ref{fig:eta_eta_residual} shows the location of the simulated microlensing events in the $\eta$ - $\eta_\text{residual}$ plane.

For each level 2 candidate we selected the object with the most nights of observation from among the candidate's objects.
The $\eta$,\ $\eta_\text{residual}$ values of these 7,457,583 best lightcurves is plotted in Figure \ref{fig:eta_eta_residual} as well.
There existed a clear distinction between the location of most of our candidates in this plane as compared to the simulated events.
This distinction was even stronger when we limited our simulated sample to events with true values of $t_E > 150$ and $\pi_E < 0.08$.
These cuts have previously been used as selection criteria for identifying microlensing events as black hole candidates \citep{Golovich2022}.
We ran a grid of proposed cuts with the criteria $\eta_\text{residual} \geq m \cdot \eta + b$ and found that $m = 3.62, b = 0.01$ retained 83.6\% of all microlensing lightcurves and 95\% of the black hole microlensing lightcurves, while removing 98.8\% of our level 2 candidates.
This left us with 92,201 candidates in our level 3 catalog. \edit{This cut could be tuned if we were interested in optimizing for a more complete sample of shorter duration events.}

\subsection{Level 4: Cleaning Data and Cutting on a Seven-Parameter Microlensing Model}
\label{sec: level 4}
\begin{figure*}[!htb]
    \centering
    \includegraphics[width=0.9\textwidth, angle=0]{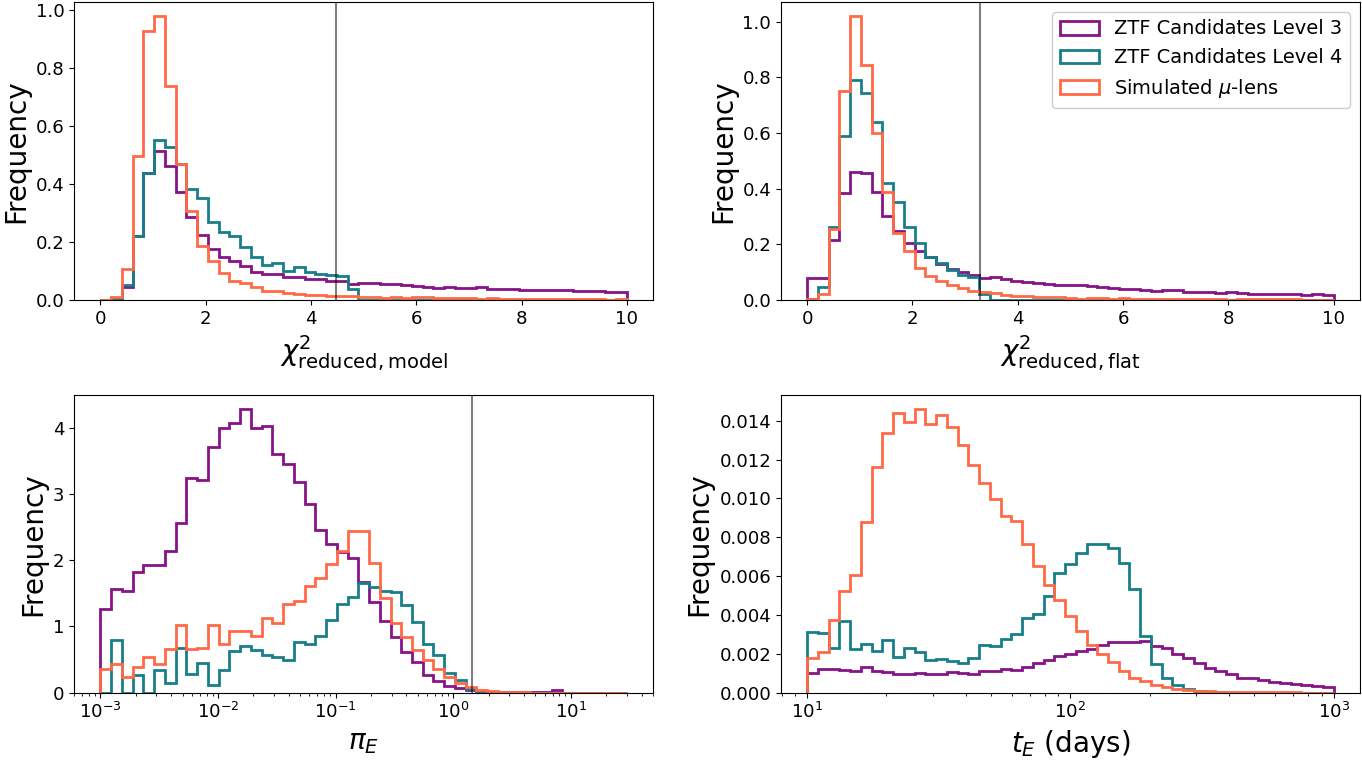}
    \caption{Fitting the seven-parameter model to both level 3 candidates (purple) and simulated microlensing lightcurves (orange) revealed cuts that removed many objects, leaving behind our level 4 catalog (blue). The $\chi^2_\text{reduced,model}$ of the entire lightcurve that was fit to a microlensing model (top-left) had a long tail in the level 3 candidates that was not present in the simulated microlensing events, motivating a cut on the $95^\text{th}$ percentile (black line). Similarly the long tail for level 3 candidates of the $\chi^2_\text{reduced,flat}$ of the lightcurve outside of $t_0 \pm 2t_E$ fit to a flat model (top-right) was not seen in the microlensing samples and was removed with a cut on the $95^\text{th}$ percentile. A small number of events with significantly larger microlensing parallaxes (bottom-left) were removed with a similar $95^\text{th}$ percentile cut. These cuts, and others, removed events with Einstein crossing times (bottom-right) longer than are detectable by our survey. \label{fig:opt_fits}}
\end{figure*}

\edit{To avoid spurious results and retain only the ``cleanest data", here we removed any lightcurve epochs with $catflag \neq 0$. We then required that all objects have at least 50 unique nights of good data with this new criteria. This brought us from 92,201 to 90,810 candidates.}
The nightly averaged magnitudes of the best lightcurve of each candidate were fit with a seven-parameter microlensing model for further analysis.
The \texttt{scipy.optimize}\texttt{.minimize} routine \citep{scipy} using Powell's method \citep{Powell1964} was used to fit each lightcurve to a point-source point-lens microlensing model including the effects of annual parallax.
\edit{The seven free parameters include $t_0$, $u_0$, $t_E$, $\pi_{E,E}$, $\pi_{E,N}$, $F_{LSN}$, $b_{SFF}$ as defined in \citep{Lu2016}.}
The average time to fit each object with this model was $1.2 \pm 0.3$ seconds.
\edit{578} candidates failed to be fit with this model and were cut.
All simulated microlensing events were also fit with this model for comparison.
The results of these fits for both populations are shown in Figure \ref{fig:opt_fits}.

Several cuts were then made on candidates using these fits.
Our first cut was to remove candidates with excessively large $\chi^2_\text{reduced,model}$ values because they could not be well fit by our microlensing model. \edit{We call this $\chi^2_\text{reduced, opt}$ in Table \ref{tab:pipeline_cuts}, where opt is an abbreviation for \texttt{scipy.optimize}, to distinguish it from the model fit in Section \ref{sec: level 4.5}.}
The threshold of $\chi^2_\text{reduced,model}=4.805$ was set by calculating the $95^\text{th}$ percentile of the simulated data and removed \edit{40,856} candidates.
We next required that the candidate's selected object had a range of epochs observed equal to at least $2 t_E$ nights outside of the event ($t_0 \pm 2t_E$), removing another \edit{34,522} candidates.
$\chi^2_\text{reduced,flat}$ was determined by calculating the $\chi^2_\text{reduced}$ comparing the brightness to the average magnitude in the region outside of the fit event for each lightcurve.
This statistic and its $95^\text{th}$ percentile value were also calculated on the simulated events and candidates above this percentile were removed.
A cut was performed on candidates above the $95^\text{th}$ percentile of $\pi_E$ to eliminate candidates where excessive parallax was producing variability in the data.
Lastly a cut was applied to $t_0$, limiting candidates to only those that peaked at least 1 $t_E$ into the survey data, removing candidates that peaked too early to be well constrained.
The combination of these cuts resulted in a list of \edit{8,649}  candidates that were either completed or ongoing events in our level 4 catalog.

\subsection{Level 4.5: Cutting on a Bayesian Microlensing Model}
\label{sec: level 4.5}
\edit{8,646 of the 8,649 candidates were successfully fit within 12 hours with a nested sampling algorithm to a point-source point-lens microlensing model with annual parallax locked to the sky location of the candidate.
We performed this more sophisticated fit to obtain error bars on our microlensing parameter measurements.
Gaussian processes were also included in the fitter to model correlated instrument noise using the procedure outlined in \citet{Golovich2022}.
This fitter fit the three most observed objects within the candidate with at least 20 nights of data to the same model simultaneously.
This model had five parameters shared by all lightcurves ($t_0, t_E, u_0, \pi_\text{E,E}, \pi_\text{E,N}$), two photometric parameters fit for each lightcurve ($m_\text{base}, b_\text{sff}$), and four Gaussian process parameters for each lightcurve ($\sigma, \rho, \omega_0, S_0$).
Therefore each candidate was fit with either 11, 17, or 23 parameters if it contained 1, 2 or 3 objects with at least 20 nights of data.
The median time to complete each of these fits was 101, 469, and 1,812 seconds for
1, 2 or 3 lightcurves respectively.
If more than three lightcurves were in the candidate (due to multiple sources with up to three filters per source) then the three lightcurves with the most nights of data were selected due to computational constraints.}

\edit{We then cut on the quality of the fit by requiring $\chi^2_\textrm{reduced} \leq 3.0$ for the data evaluated against the model fit. We also required that candidates have observations on the rising side of the lightcurve ($t_0 - t_E \geq 58194$); this is a re-implementation of an earlier cut with this more rigorous fit. We have 5,496 remaining candidates after these cuts. We waited to make the rest of the quality cuts before separating out the ongoing events since they may have poorly constrained parameters.}

\subsection{Level Ongoing}
\label{sec: level ongoing}
All of the \edit{level 4.5} candidates are interesting objects worthy of further investigation.
We divided our sample into candidates with peaks during the survey ($t_0 + t_E \leq 59243$) and \edit{ongoing} candidates with peaks after the survey was completed ($t_0 + t_E > 59243$).
These sub-samples of \edit{3,938 peaked and 1,558 ongoing} candidates respectively serve two different purposes.
Calculating the statistical properties of microlensing candidates within ZTF data requires a clean sample of completed events.
We therefore sought to further remove contaminants from the \edit{3,938} candidates that had peaked within the survey \edit{(see Section \ref{sec: level 5})}.
The \edit{1,558} candidates that had yet to peak need to be further monitored while they are rising in brightness until they reach their peak amplification.
At this point they could be fit with a microlensing model to determine whether they are good events and perhaps even black hole candidates worthy of astrometric followup.

\subsection{Level 5: Cutting on Completed Events}
\label{sec: level 5}

Data quality cuts were applied to the resulting fits, as outlined in Table \ref{tab:pipeline_cuts}.
The fractional error on $t_E$ was required to be below 20\% to ensure that a final distribution of Einstein crossing times would be well defined.
Cuts on $|u_0| \leq 1.0$ and $b_\text{sff} \leq \edit{1.5}$ are commonly accepted limits for the selection of high quality events.

The candidate was required to have $4t_E$ of observations that were outside of significant magnification ($t_0 \pm 2t_E$), \edit{which is a re-implementation of a previous cut with this more rigorous fit}.
All of these cuts reduced the sample down to \edit{950} candidates in our level 5 catalog.
We additionally fit the \edit{1,558} candidate to the same nested Bayesian sampler but with only the one single lightcurve containing the most nights of observations.
This gave us the error on the $t_0$ and $t_E$ measurements that could be used to determine which of these candidates should be astrometrically observed.

\subsection{Level 6: Manual Lightcurve Inspection and Crossreference}
\label{sec: level 6}
\begin{figure*}[!htb]
    \centering
    \includegraphics[width=0.9\textwidth, angle=0]{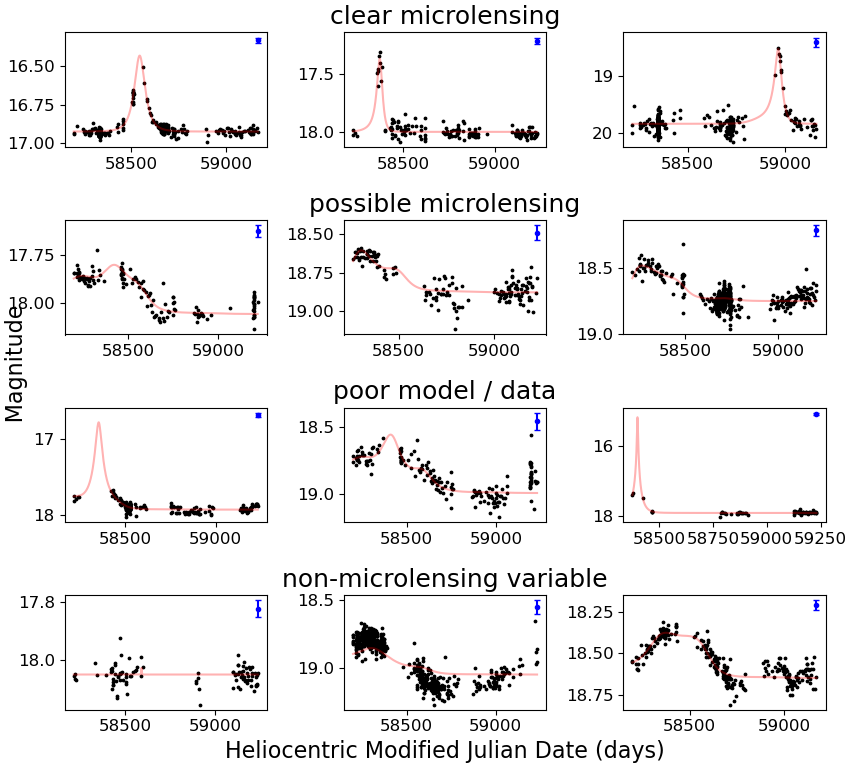}
    \caption{Lightcurves exemplifying the four labels manually assigned to level 5 candidates. Only the clear microlensing events were selected for our final level 6 catalog. \edit{The blue point in the upper right corner for each lightcurve represents the average error for each lightcurve.} The limitations of our method can be seen in the persistence of non-microlensing variables that had additional variation not explained by microlensing, as well as lightcurves where the model was not supported by the data labelled as poor model / data. The possible microlensing events had insufficient data to be confidently named microlensing. Those appearing to rise in brightness could be confirmed with further observations. \label{fig:label_examples}}
\end{figure*}

There were \edit{950} candidates remaining in the level 5 catalog.
There were several failure modes of our pipeline that could only be addressed by manually labeling each of these candidates.
To facilitate this, we constructed a website that provided access to the candidate information contained within the \texttt{PUZLE} database tables and lightcurve plots generated by \texttt{zort}.
This website \edit{is} a docker container running a flask application that was served on NERSC's Spin platform.
The website connected inspectors to the NERSC databases and on-site storage.
Each of the candidate labels was derived after inspecting the data by eye and grouping the candidates into common categories.
The description and final number of candidates with each label after manual inspection are outlined in Table \ref{tab:dr5_labels}.
Examples of each candidates with each label are shown in Figure \ref{fig:label_examples}.

\begin{table*}[t]
  \centering
    \begin{tabular}{p{0.2\linewidth}|p{0.6\linewidth}|p{0.1\linewidth}}
      \hline
      \hline
      Label & Description & Number \\
      \hline
      clear microlensing & Model accurately follows a rise and fall in brightness with a clear region of uncorrelated non-lensing brightness measured outside of the event. & \edit{60} \\
      possible \newline microlensing & Model accurately follows either a rise or fall in brightness, does not have sufficient area of non-lensing outside of the event, or has fewer than 10 points deviating from baseline. & \edit{580} \\
      poor model / data & Model predicts a significant variation in brightness in areas without sufficient data. & \edit{229} \\
      non-microlensing variable & Correlated deviation from the model is present in either the non-lensing region outside of the event, or the lightcurve is similar in appearance to a supernova. & \edit{81} \\
    \end{tabular}
    \caption{Microlensing Event Visual Inspection Labels}
    \label{tab:dr5_labels}
\end{table*}

\begin{figure*}[!htb]
    \centering
    \includegraphics[width=0.9\textwidth, angle=0]{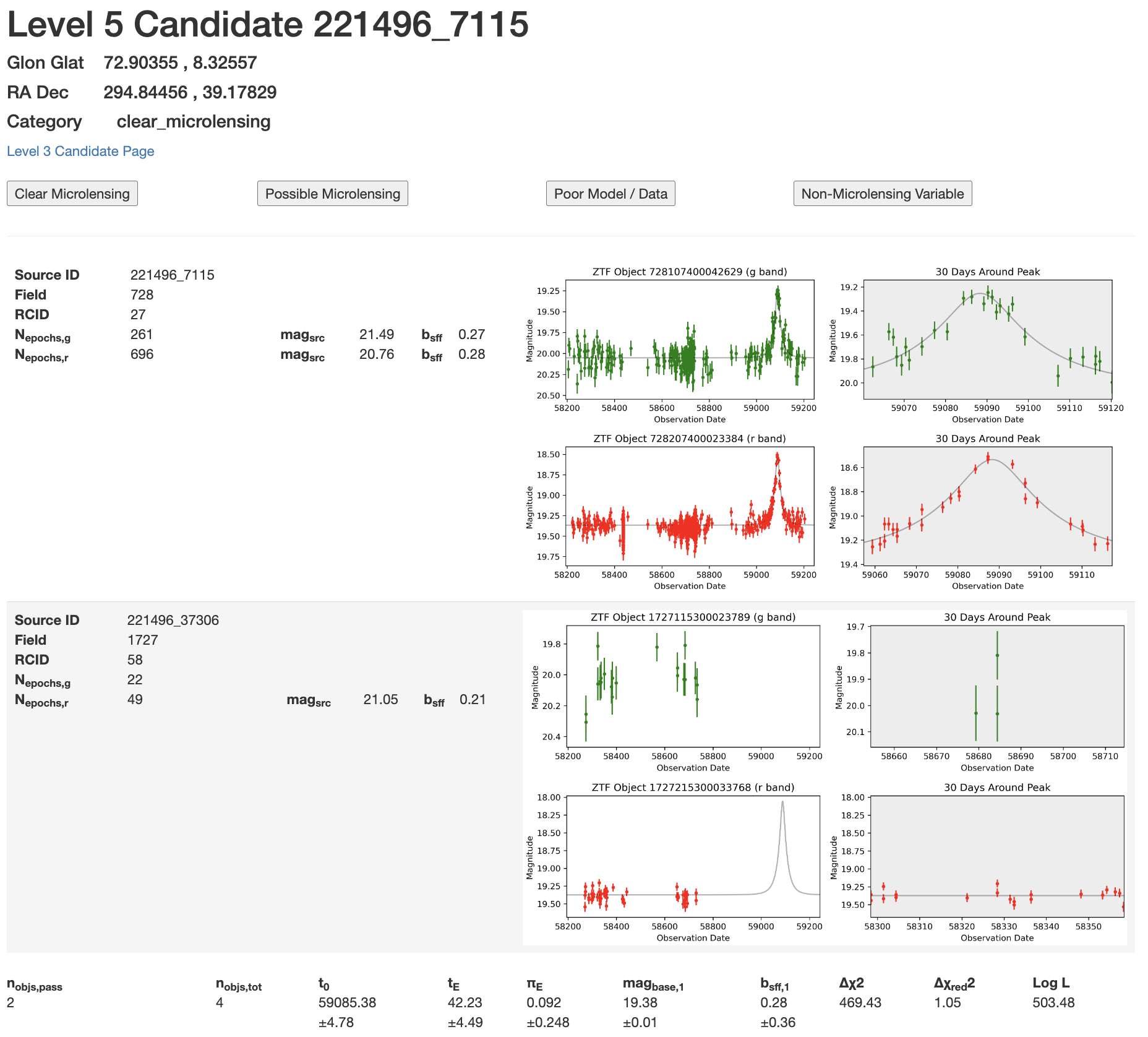}
    \caption{Level 5 candidates were manually screened by a human expert on the \texttt{PUZLE} website. Inspectors were presented with all objects in the candidate star grouped by source. Model curves were plotted on those lightcurves that were included in the Bayesian fit. The maximum a posteriori probability and 1$\sigma$ error bars of Bayesian fit parameters were also shown. The inspector then assigned one of four labels to the candidate.  \label{fig:labeling_example}}
\end{figure*}

The inspector was shown a display page with the four labels displayed above the \texttt{zort} lightcurves alongside information from the database.
The model derived from the Bayesian fit was plotted onto the 1 to 3 objects that contributed to fitting the model and was absent from those objects within the candidate that did not.
The user then selected which label best matched the candidate.
The user either selected a scoring mode, where the page was automatically advanced to another unlabelled candidate, or a view mode, where the page remained on the candidate after selecting a label.
An example page from the labelling process is shown in Figure \ref{fig:labeling_example}. \edit{66 candidates were assigned the clear microlensing label.}

\edit{After visual inspection, each of the candidates was checked in SIMBAD to identify if they have some other known source of variability. 5 objects were identified as quasars in SDSS \citep[which is expected to be 99.8\% complete][]{Lyke:2020} and 1 was identified as a Be star with significant variability \citep{Pye:1983}. The summary of the final 60 candidate events can be found in Table \ref{tab:events}.}

\section{Results} \label{sec:results}

\begin{figure*}[!htb]
    \centering
    \includegraphics[width=0.9\textwidth, angle=0]{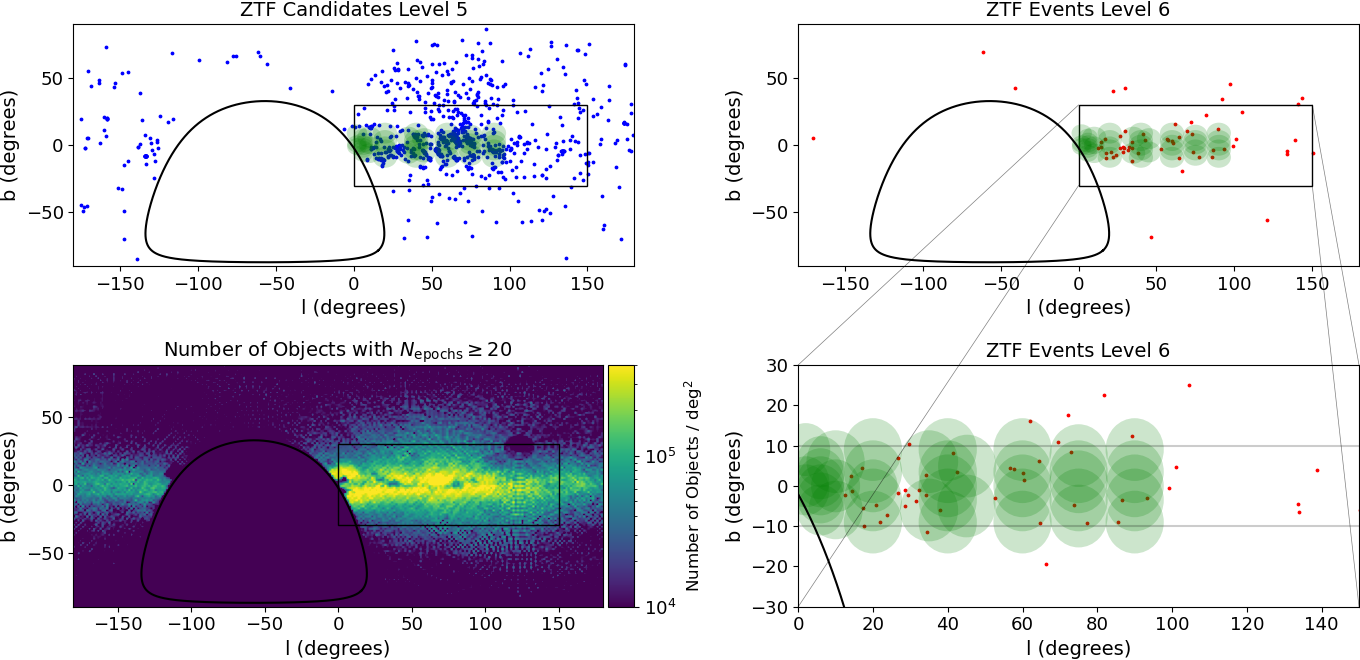}
    \caption{Identifying our final level 6 candidates by manual inspection produced a list of clear microlensing events scattered throughout the night sky. The \edit{950} level 5 candidates (blue) were positioned throughout the sky (top-left), while the \edit{60} level 6 events (red, top-right) are primarily located within the Galactic plane (bottom-right). \edit{35 of the 60 level 6 events (58\%)} appear within the area of the \texttt{PopSyCLE} simulations we performed (footprints in green). However, there are a significant number of events at larger Galactic latitudes. This can be explained by the relatively larger number of objects with $N_\text{epochs} \geq 20$ \edit{(with catflag $<$ 32768)} observed at these locations in DR5 (bottom-left), contamination by long-duration variables, or perhaps the presence of MACHOs in the stellar halo. \label{fig:level5_cands_on_sky}}
\end{figure*}

\begin{figure*}[!htb]
    \centering
    \includegraphics[width=0.9\textwidth, angle=0]{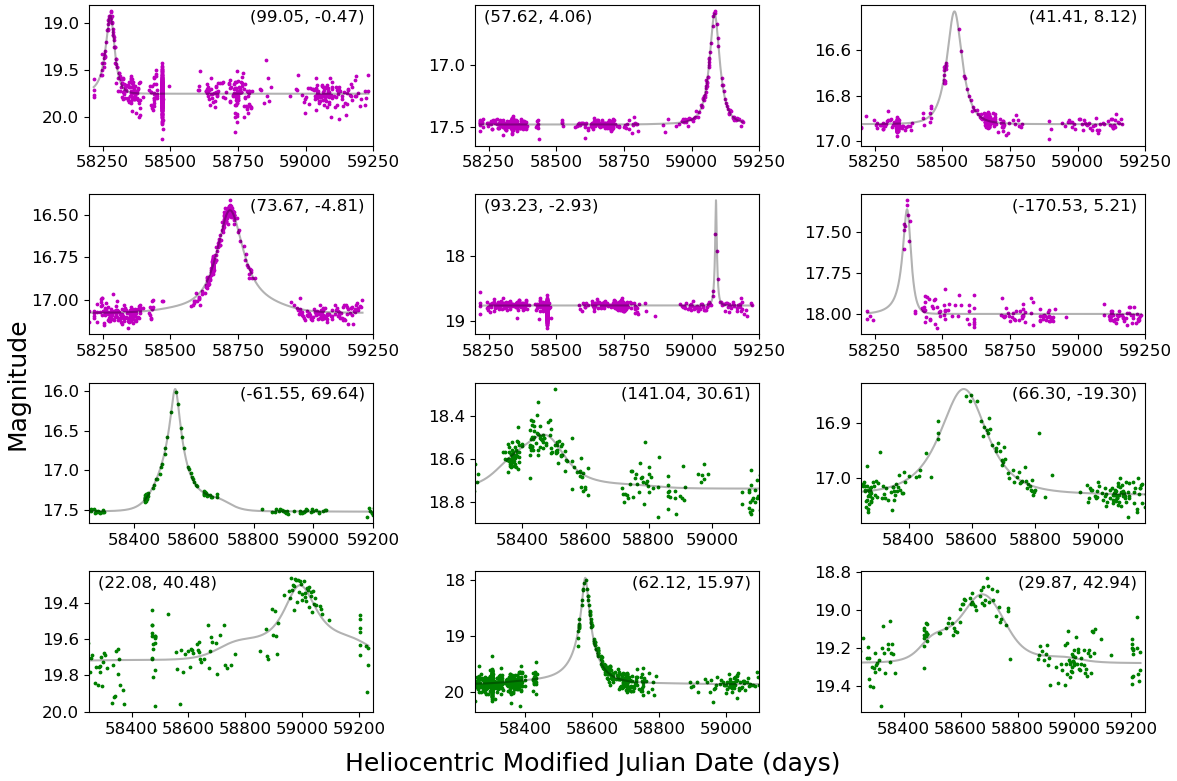}
    \caption{Our level 6 catalog contains \edit{41} Galactic plane microlensing events (purple) and \edit{19} events outside of the plane (green) \edit{ordered by increasing $\sigma_{t_E}/t_E$}. These sample lightcurves show the variety of timescales and magnitudes identified by our pipeline. The Galactic longitude and latitude of each event is printed in each corner. These \edit{19} events nearly \edit{double} the total number of microlensing events yet discovered outside of the Galactic plane and bulge.  \label{fig:level6_cands_lightcurves}}
\end{figure*}

Our final catalog of level 6 candidates contains \edit{60} clear microlensing events. The sky location of these events are shown in Figure \ref{fig:level5_cands_on_sky}.
\edit{41 (68\%)} of the events are within the Galactic plane ($|b| \leq 15^\circ$) and \edit{19 events (32\%}) are outside of it.
The distribution of events roughly follows the density of observable objects within DR5, also included in the figure.
This indicates that our pipeline is finding microlensing events proportional to the stellar density in each part of the sky.
A selected number of these events can be seen, divided by sky location, in Figure \ref{fig:level6_cands_lightcurves}.
There are a large number of events located in parts of the sky far outside of the Galactic plane.
Microlensing from two spatially coincident stars is far less likely in these regions.
We further discuss the possible origin of these events in Section \ref{sec:discussion}.
All \edit{60} microlensing events have been posted online for public use \citep{puzle_microlensing_catalogs}, following the schema outlined in Table \ref{tab:exported_data_schema}.

\begin{figure*}[!htb]
    \centering
    \includegraphics[width=0.9\textwidth, angle=0]{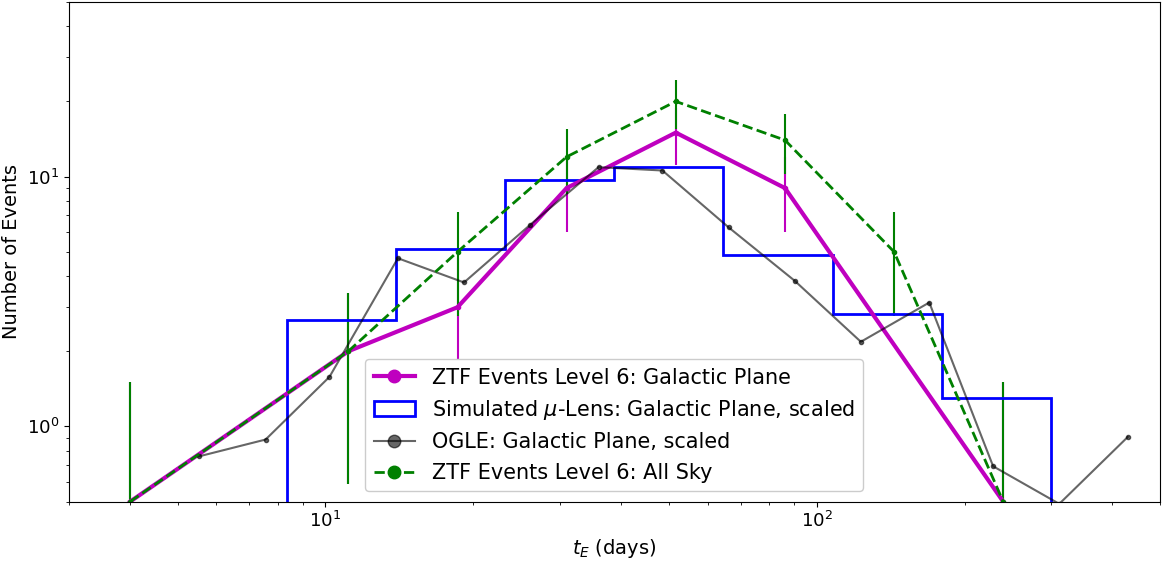}
    \caption{The Einstein crossing times for the level 6 event catalog across the entire sky (purple) and those found within our simulated \texttt{PopSyCLE} Galactic plane fields (green) appear similar. Each curve peaks at around 50 days. 
    These peaks are at longer crossing times than expected from the observable events contained within the overlapping \texttt{PopSyCLE} simulations (blue) scaled to match the number of ZTF Galactic plane level 6 events. The OGLE-IV Galactic plane fields (black), also scaled to match these events, largely agrees with our simulations. However, it also shows the limits of our small sample size and short-duration survey to be able to detect the longest duration events.\label{fig:level5_cands_tE}}
\end{figure*}

The distribution of Einstein crossing times in our level 6 events are shown in Figure \ref{fig:level5_cands_tE}.
Short duration events as short as \edit{one day} are recorded, increasing in number up to a peak at $t_E \approx 50$ days.
At longer durations there is a
decline with no events found at $t_E \geq \edit{120}$ days.
Several cuts within our pipeline require significant duration of observations outside of the microlensing event.
These constraints, combined with the limited duration of DR5, place a strong upper limit on the timescale of events we can detect.
We compare this distribution \edit{to} previous simulations and other surveys in Section \ref{sec:discussion}.

\begin{figure*}[!htb]
    \centering
    \includegraphics[width=0.9\textwidth, angle=0]{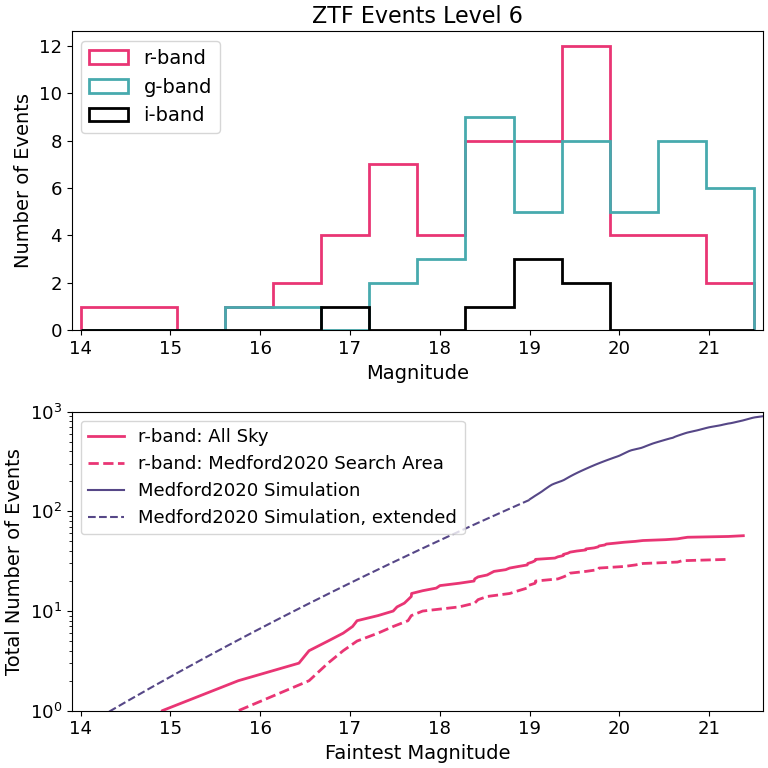}
    \caption{The baseline magnitudes of level 6 events are consistent across filters and follow the shape of the luminosity function predicted by \citet{Medford2020b} up to $19^\text{th}$ magnitude. The \edit{60} level 6 events contain \edit{58} r-band, \edit{51} g-band and \edit{7} i-band baseline magnitudes that met the requirements to be fit by the Bayesian fitter. The distribution of these baseline magnitudes (top) appears increasingly plentiful at bright magnitudes, with all three filters dropping off to lower numbers past $19^\text{th}$ magnitude. Comparing the cumulative number of events with r-band baseline magnitudes brighter than a given magnitude (bottom, red) to the \citet{Medford2020b} 3-year $l \geq 10^\circ$ predictions (purple) show a similar shape for magnitudes brighter than 19$^\text{th}$, but with a clear deficit of events due to the selection effects of our pipeline. At r-band magnitudes fainter than 19$^\text{th}$ we are increasingly unable to observe any events. This suggests that executing a search on ZTF data reprocessed to combine images into co-additions could yield many more events as predicted by \citet{Medford2020b}. \label{fig:level6_magniutdes}}
\end{figure*}

The luminosity function of our level 6 events is shown in Figure \ref{fig:level6_magniutdes}.
The magnitude shown is the median baseline magnitude, as determined by our Bayesian fitter, for all lightcurves of the same filter belonging to an event.
\edit{58} of the events contain at least one lightcurve that was fit by the Bayesian fitter in the r-band, \edit{51} of the events have fit data in g-band, and \edit{7} events possess fit i-band.
All three filters peak in number at $19^\text{th}$ magnitude, followed by a sharp decline.
Our pipeline appears to be removing events fainter than $19^\text{th}$ magnitude.
The baseline magnitudes for g-band events also decline but at a slower rate.
We compare the r-band luminosity function to simulation in Section \ref{sec:discussion}.

\begin{figure*}[!htb]
    \centering
    \includegraphics[width=0.9\textwidth, angle=0]{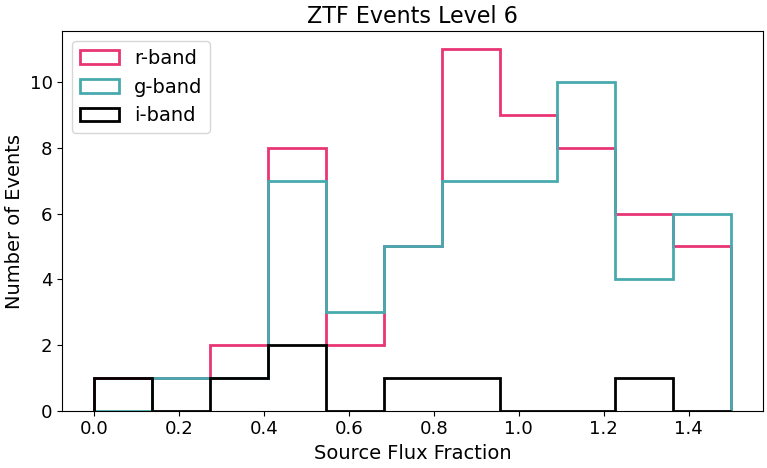}
    \caption{The source flux fractions of level 6 events drawn from across the sky are more uniform than results drawn from other works aimed toward the Galactic bulge. While the \edit{58} r-band, \edit{51} g-band and \edit{7} i-band baseline magnitudes that met the requirements to be fit by the Bayesian fitter each have slightly different distributions, they all have very few events with low source flux fractions ($b_\text{sff} \lesssim 0.2$). This is expected in the relatively low stellar densities of the outer Galactic plane where the presence of neighboring stars is less prevalent than in the Galactic bulge. The g-band and r-band magnitudes peak at around $b_\text{sff} = 1$, where nearly all of the event flux originates in the source. The presence of sources with $b_\text{sff} > 1$ indicates that the background is overestimated, indicating the accuracy limits of the DR5 photometric catalogs. \label{fig:level6_source_flux_fraction}}
\end{figure*}

\citet{Medford2020b} predicted that events observed throughout the outer Galactic plane would have larger source flux fractions ($b_\text{sff}$) than events observed in the Galactic bulge.
Figure \ref{fig:level6_source_flux_fraction} shows that our level 6 candidates have larger source flux fractions in line with this prediction.
Very few events are seen with $b_\text{sff} \leq 0.2$ as the relatively small stellar densities of the outer Galactic bulge and the stellar halo prevent many neighboring stars from appearing in the instrument's observational aperture.
This is in contrast to the findings of most Galactic bulge surveys that find a bi-modal distribution with a peak around $b_\text{sff} \lesssim 0.2$. \edit{It is also possible that we may be biased against some high blend events due to the smaller observed change in magnitude.}
Both g-band and r-band objects have an increasing number of events at larger source flux fractions and peak at approximately $b_\text{sff} \approx 1$.
However, both bands and particularly the g-band shows an excess of events at $b_\text{sff} \geq 1$.
This indicates that there is an overestimate of the background noise present in the photometric measurements.
Reprocessing the ZTF observations at the location of these events with calibration parameters tuned to the particulars of each event's fields could improve the estimates of these backgrounds and reduce the number of events with excessive source flux fractions.

\section{Discussion} \label{sec:discussion}

Throughout this work we have developed new techniques for efficiently finding events in the massive ZTF dataset.
Simulations in the same areas of the sky give us expectations against which to compare our results.
We also can compare our results to surveys designed specifically for microlensing to get a handle on how well our non-traditional methods have performed.

Our catalog of microlensing events contains objects in regions of the sky where previous microlensing campaigns have not observed.
We therefore need to limit our level 6 catalog to a sub-sample that overlaps with previous OGLE observations and \texttt{PopSyCLE} simulations to be able to make comparisons \edit{between $t_E$, $\pi_E$ $b_{sff}$, and baseline magnitude}.
The locations of our \texttt{PopSyCLE} microlensing simulations are shown in the right panels of Figure \ref{fig:level5_cands_on_sky}.
For this comparison we select observable \texttt{PopSyCLE} events that are required to have $u_0 \leq 1.0$, a maximum amplification $\Delta m_r \geq 0.3$ and a baseline magnitude $m_r \leq 21.5$.
These requirements produce events that are observable to ZTF, albeit without a correction for cadence or time span.
We then scale the distribution of Einstein crossing times of these events to match the total number of events in our Galactic plane sub-sample for comparison in Figure \ref{fig:level5_cands_tE}.

The short-duration slope of our Galactic plane sub-sample is approximately equal to the simulation.
Both our sub-sample and the \texttt{PopSyCLE} catalogs peak at approximately 50 days. \edit{Our sub-sample mostly agrees above this peak with a potential small excess of long-duration events.}
Increased ZTF survey time could more than double the microlensing event sample size and \edit{either confirm agreement or an excess compared to the simulation.}
At the longest durations our pipeline is unable to recover events.
The cut on sufficient baseline applied to level 4 candidates required $4t_E$ of observations outside of the event ($t_0 \pm 2t_E$).
Applying this requirement for $8t_E$ of data to a three year survey puts an upper limit on our data of approximately $t_E \sim 135$ days.
This explains the sharp drop-off in long-duration events around this point.

\citet{Mroz2020A} conducted the largest Galactic plane search for microlensing events prior to this work, discovering 630 events in the OGLE survey observing 3000 square degrees of the Galactic plane between 2013 and 2019.
ZTF observes an even larger footprint of the Galactic plane, with 6900 square degrees between $-15^\circ \leq b \leq 15^\circ$ and either $10^\circ \leq l \leq 180^\circ$ or $-180^\circ \leq l \leq -120^\circ$.
However, many of these fields point toward or are near the Galactic anti-center where stellar densities are low compared to the Galactic bulge fields observed by \edit{OGLE}.
OGLE also observed their fields for three more years than the ZTF DR5 dataset.
We include in Figure \ref{fig:level5_cands_tE} the distribution of Einstein crossing times found by \citet{Mroz2020A} for the Galactic plane $(|l| \geq 20^\circ)$, re-scaled to match the number of events in our sample.
The \texttt{PopSyCLE} and OGLE distributions are in good agreement and predict a relatively equal number of short and long duration events around a peak of approximately 50 days.
While the \texttt{PopSyCLE} fields are run at different Galactic longitudes than the OGLE fields, this is in line with the findings of \citet{Mroz2020A} that the Einstein crossing time distribution within the plane is independent of longitude for $|l| \geq 20^\circ$.

\begin{figure*}[!htb]
    \centering
    \includegraphics[width=0.9\textwidth, angle=0]{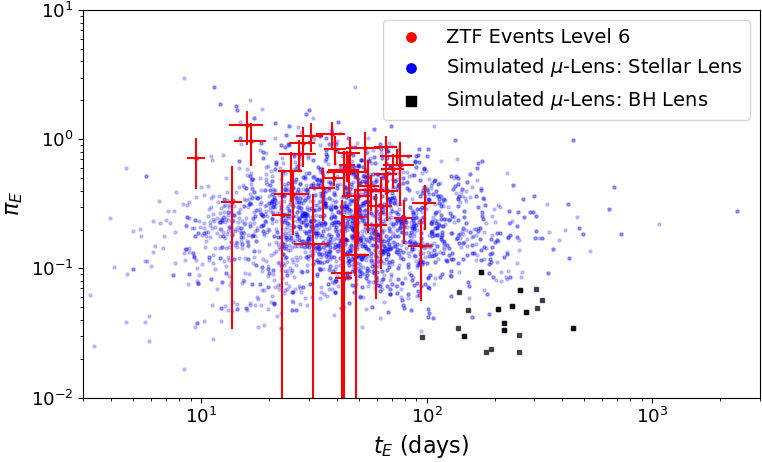}
    \caption{The Einstein crossing times and microlensing parallax of all level 6 events with $1\sigma$ errors (red) fall right on top of the region where the \texttt{PopSyCLE} catalogs contain events with stellar lenses (blue). We do not find any events within the region of $t_E-\pi_E$ space where \texttt{PopSyCLE} predicts black hole lenses (black). The quality of our data prevents us from constraining small values of $\pi_E$, and the duration of our survey prevents the detection of the longest duration events. The \texttt{PopSyCLE} events shown are all of the observable events within the simulations. \label{fig:level5_cands_tE_piE}}
\end{figure*}

\citet{Lam2020} found that black holes could be identified by placing microlensing events in the $t_E$-$\pi_E$ space and looking for the events with the \edit{longest} $t_E$ and the \edit{smallest} $\pi_E$.
In Figure \ref{fig:level5_cands_tE_piE} we place our events in this space, alongside all observable events from within our \texttt{PopSyCLE} simulations.
The simulated events are separated between those \edit{with a} stellar lens and those with a black hole lens.
Level 6 events appear within parameter space most often occupied by events with stellar lenses.
This can be caused by several factors.
It is possible that all of our events are caused by stellar lenses.
The relatively short timescale of DR5 combined with our strict requirements for data outside $2t_E$ of our event also places a limit on the longest events we are able to detect.
And finally, small values of $\pi_E$ are difficult to constrain with the quality of photometry that ZTF produces.
The result is that we are unable to confidently claim that any of our microlensing events are black hole candidates.

\citet{Medford2020b} produced an estimate for the number of events detectable as a function of the faintest observable magnitude of ZTF imagery.
They predicted that ZTF would discover $\sim$500 events in the outer Galaxy for a r-band limiting magnitude of 20.6, and $\sim$135 events in the same region for a r-band limiting magnitude of 19.
In Figure \ref{fig:level6_magniutdes} we compare the cumulative distribution function of our luminosity function to these results.
\edit{\citet{Medford2020b} has overpredicted the number of microlensing events discovered by the \texttt{PUZLE} pipeline.}
The sub-sample of level 6 events that fall within the \citet{Medford2020b} estimate footprint ($10^\circ \leq l \leq 100^\circ$, $-10^\circ \leq b \leq 10^\circ$) has a similar shape to the predictions.
However, we detect only \edit{35} microlensing events within the \citet{Medford2020b} estimate footprint, \edit{19} of which have $m_\text{base,r} \leq 19$.

We attribute the gap between the predictions of \citet{Medford2020b} and our estimate footprint sub-sample to several factors.
The cut on level 2 candidates removed 16.4\% of simulated microlensing events, and the three threshold cuts on level 3 candidates each removed 5\% of \edit{the remaining} simulated events.
This accounts for missing 28\% of possibly observable events.
The level 4 catalog are all candidates that are well fit by a microlensing model and the cuts applied to them are to get a sample of high quality events.
Therefore, the order of magnitude drop from \edit{5,496} candidates to \edit{950} candidates likely removes many true events.
Strict requirements on the amount of observable data outside of the event removes long duration events from the final catalog.

The level 2 and level 3 cuts retained 72\% of possible observable events. \edit{We selected a random 100 lightcurves from level 4.5 and visually inspected them and found 2\% that would have been characterized as ``clear microlensing," if they had passed to level 5.}
The \edit{60} events we discovered \edit{are} \edit{12}\% of the $\sim$500 predicted by \citet{Medford2020b}.
We would have to assume that the \edit{cuts between level 3 and 4}  removed an additional \edit{81}\% of true events from our catalog to match these predictions.
While this is possible, given that the \edit{cuts between level 3 and 4} selected only the highest quality events from a list of possible microlensing candidates, the difference between our yields and the predictions suggests incompleteness to our method.
\citet{Medford2020b} did not take into account the effect of gaps in the data due to either weather or seasonality. \edit{It also did not cut out events that did not have enough baseline and only required that the peak be in the survey rather than $t_0 \pm t_E$ being in the survey, which may cut out many true events.}
This makes the \edit{discovered} number of microlensing events lower than their prediction.

The lack of fainter sources in the level 6 catalog indicates that the ZTF limiting magnitude of 20.6 used in \citet{Medford2020b} may have been too large.
This result also suggests that our pipeline selectively removed fainter stars with relatively larger photometric errors.
This finding also points to the number of additional microlensing events that could be discovered if ZTF were reprocessed by combining subsequent observations into co-additions.
\citet{Medford2020b} show the non-linear gains that are predicted to be achieve\edit{d} by this method, and our results corroborate that claim.

\edit{\cite{Mroz2020B} and \cite{Rodriguez:2022} conducted similar surveys of ZTF data for microlensing events (DR2 and through DR5 respectively). They found 30 and 60 events, respectively, 12 and 28 of which we also recovered. The 12 from \cite{Mroz2020B} are also recovered in \cite{Rodriguez:2022}, so we discovered 32 new events. See Appendix \ref{Appendix: lit comparison} for a more detailed comparison.}

\subsection{Black Hole Candidates}

The level 6 catalog is not ideal for searching for black hole candidates due to the limit on $t_E$ that our cuts impose.
However, there are numerous microlensing events within ZTF-I that are still on-going.
These were cut in level 4 due to the requirement that all candidates have a time of closest approach within $1t_E$ of the end of DR5 observations.
This requirement limits our level 5 and 6 samples to completed events.
\edit{1,558} candidates that had passed all level \edit{4.5} cuts \edit{and $t_0 + t_E > 59243$ MJD were classified as} ``level ongoing''.
If black hole microlensing events are contained within ZTF observations, they would most likely be long duration and therefore still ongoing.

\begin{figure*}[!htb]
    \centering
    \includegraphics[width=0.9\textwidth, angle=0]{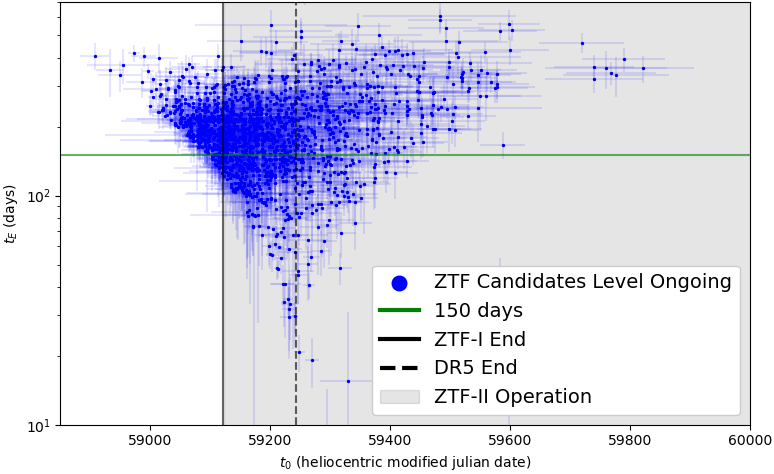}
    \caption{There are \edit{1,558} candidates that passed all cuts up to level \edit{4.5 and are} still ongoing as of the end of the DR5 dataset. \edit{They are labeled ``level ongoing" in Table \ref{tab:pipeline_cuts}.} These candidates have a time of closest approach ($t_0$) within one Einstein crossing time ($t_E$) of the end of DR5 observations (black solid line) or later. Events with an Einstein crossing time greater than 150 days (green line) that are continuing throughout ZTF-II (black shaded) are good candidates for continued observation. If they are found to decline in brightness in agreement with a microlensing model, they may be followed-up astrometrically to search for black hole lenses. \label{fig:level4_cands_ongoing_t0_tE}}
\end{figure*}

The Einstein crossing times versus times of closest approach of level ongoing candidates are shown in Figure \ref{fig:level4_cands_ongoing_t0_tE}.
The majority are modeled to have already passed peak brightness by the end of DR5 data.
However, \edit{802} candidates have $t_E \geq 150$ days and are projected to hit peak brightness during the ZTF-II campaign.
These sources could be continuously observed to see if any decline in brightness is in agreement with a microlensing model.
If found to do so, astrometric follow up could be combined with photometric measurement to weigh the mass of the lens and possibly make a detection of an isolated black hole.
All \edit{1,558} level ongoing candidates have been posted online for public use \citep{puzle_microlensing_catalogs}, following the schema outlined in Table \ref{tab:exported_data_schema}.

\begin{table*}[t]
  \centering
    \begin{tabular}{ccc}
      \hline
      \hline
      Column Name & Datatype & Unit \\
      \hline
      id & string & - \\
      ra & float & degrees \\
      dec & float & degrees \\
      t0 & float & hmjd \\
      t0\_err & float & hmjd \\
      tE & float & days \\
      tE\_err & float & days \\
      u0\_amp & float & - \\
      u0\_amp\_err & float & - \\
      piE\_E & float & - \\
      piE\_E\_err & float & - \\
      piE\_N & float & - \\
      piE\_N\_err & float & - \\
      mag\_base\_r & float & magnitude \\
      mag\_base\_err\_r & float & magnitude \\
      mag\_base\_g & float & magnitude \\
      mag\_base\_err\_g & float & magnitude \\
      mag\_base\_i & float & magnitude \\
      mag\_base\_err\_i & float & magnitude \\
      b\_sff\_r & float & - \\
      b\_sff\_err\_r & float & - \\
      b\_sff\_g & float & - \\
      b\_sff\_err\_g & float & - \\
      b\_sff\_i & float & - \\
      b\_sff\_err\_i & float & - \\
    \end{tabular}
    \caption{Exported Candidate Schema. Exported data have been posted online for public use \citep{puzle_microlensing_catalogs}.}
    \label{tab:exported_data_schema}
\end{table*}

Our search for microlensing in the public data releases was not performed in real-time.
This approach is limited when attempting to find targets for astrometric follow-up.
However, all of our microlensing event selection methods have been designed to be computationally inexpensive and quick to execute.

\subsection{Events Outside the Galactic Plane}

Our final level 6 catalog contains \edit{19} events outside of the Galactic plane ($|b| \geq 15^\circ$), with several examples shown in Figure \ref{fig:level6_cands_lightcurves}.
Previously, attempts to detect microlensing with lenses in the stellar halo by observing background stars in the Magellanic Clouds and the Andromeda \edit{G}alaxy have yielded only a few events.
In \citet{Wyrzykowski2011a} and \citet{Wyrzykowski2011b}, the OGLE survey discovered 2 and 3 events over 8 years toward the Large Magellanic Clouds and Small Magellanic Clouds respectively.
\citet{Alcock2000} discovered 13 to 17 events in the 5.7 years that the MACHO collaboration observed the Large Magellanic Clouds.
\citet{Tisserand2007} detected 1 event in both the Magellanic Clouds in the 6.7 years of the EROS-2 survey.
Surveys toward the Andromeda \edit{G}alaxy have detected even fewer events, with \citet{Calchi_Novati_2014} detecting 3 events over the 4 year PLAN campaign and \citet{Niikura_2019} discovering 1 short duration event after 7 hours of dense sampling.
Our pipeline's \edit{19} events \edit{are} more than all 20 to 25 previously discovered events, nearly \edit{doubling} the total number of microlensing events discovered outside of the Galactic plane and bulge.

There are several explanations for the presence of microlensing events in significant numbers outside of the Galactic plane in our catalog.
The probability of two stars appearing to co-align in lower stellar densities is small, but such yields could be possible by searching across ZTF's all-sky footprint.
Our \edit{level 4} sample could be contaminated by variations in stellar brightness that are well fit by a microlensing model but are instead other long-duration variability.
\edit{The contamination is difficult to quantify at this level, as there is simply not enough data to disambiguate the scenarios.
However, the level 5 and 6 catalogs should be free of contamination from stellar variability.}
The most exotic explanation would be the presence of MACHOs that exist in large enough numbers to lens background stars.
While previous works claim to have eliminated the possibility of large numbers of MACHOs \citep{Alcock2001, Wyrzykowski2011b}, more recent simulations that replace a monochromatic mass with an extended mass distribution have re-opened the door for MACHOs \citep{Carr2016, Calcino2018}.
Our detections of events toward the stellar halo could be interpreted as evidence for these claims, although further validation is needed.

\subsection{Future Improvements}

There are several aspects of our pipeline that could be improved to increase the accuracy and yield of our results.
Our DR3 to DR5 conversion method leaves out additional i-band observations that wouldn't have been seen in earlier versions.
A future implementation could keep an ordered set that tracks the \texttt{object} IDs that have been used and then uses each \texttt{radec\_map} to push the new objects into an existing \texttt{source}.
 \edit{We could also do another check of how well the microlensing model fits the lightcurve after fitting more robust models. This could reduce the presence of events with large baseline variability which were classified as ``non-microlensing variables".}
 \edit{We could add a cut to confirm that the peak of the model is covered by the data, as many lightcurves classified as having ``poor model/data" did not have peak coverage.}
Simulated microlensing lightcurves could have been injected into our sample at the beginning of the pipeline to better measure completeness throughout the process.
These lightcurves could also be mixed into the web portal candidates before expert scoring to measure the effect of human bias in identifying ``clear microlensing'' events.
More careful comparison to simulations after completeness correction would also more accurately assess the effectiveness of our pipeline.

Several large synoptic surveys are due to see first light in the next few years, including the Rubin Observatory's Legacy Survey of Space and Time (LSST) and the Nancy Grace Roman Space Telescope (Roman).
The techniques developed in this work are designed to accommodate the massive datasets that these surveys will generate, \edit{which will eclipse the size of the largest present-day datasets by orders of magnitude}.
Our pipeline was designed to remove large numbers of events quickly and without significant computational cost.
We fit progressively more complicated models onto the data in order to prevent spending resources fitting events that are less likely to be true microlensing.
We eliminate all but the last few candidates before requiring human intervention.
This approach, successfully executed on ZTF data, can unlock the potential for the next generation of massive all-sky surveys to become microlensing machines.
\edit{LSST will be similar to ZTF in terms of sky coverage and cadence\footnote{\url{https://docushare.lsst.org/docushare/dsweb/Get/Document-38411}}.
However, for Roman, substantial modifications will have to be made to the pipeline to account for the large temporal gaps between the Galactic bulge observing windows.}
\edit{The ultimate goal is for identification and classification of microlensing events from large surveys to be fully automated, completely free from the intervention of a human expert. This will allow us to more readily review a dataset orders of magnitude larger than in this paper, ensure reproducibility, and avoid human error.}
\edit{Using machine learning methods to accomplish this is an active area of research, especially approaching the era of the Roman Space Telescope's Galactic Bulge Time Domain Survey and the Vera C. Rubin Legacy Survey of Space and Time \citep{Wyrzykowski2015, Godines:2019, Mroz:2020, Zhang_lfi:2021, Gezer:2022}.}

\begin{acknowledgments}
\edit{We thank Antonio Rodriguez and Przemek Mr\'{o}z for alerting us to the need to use the $catflag=0$ criteria. We thank the referee for helpful comments that improved the clarity of the paper.} This work is based on observations obtained with the Samuel Oschin Telescope 48-inch and the 60-inch Telescope at the Palomar Observatory as part of the Zwicky Transient Facility project.
This research used resources of the National Energy Research Scientific Computing Center, a DOE Office of Science User Facility supported by the Office of Science of the U.S. Department of Energy.
We acknowledge support from the DOE under grant DE-AC02-05CH11231, Analytical Modeling for Extreme-Scale Computing Environments.
J.R.L. and C.Y.L. acknowledge support by the National Science Foundation under Grant No. 1909641 and the National Aeronautics and Space Administration (NASA) under contract No. NNG16PJ26C issued through the WFIRST (now Roman) Science Investigation Teams Program.
\end{acknowledgments}

\bibliographystyle{aasjournal}
\bibliography{references.bib}

\begin{thebibliography}{}
\expandafter\ifx\csname natexlab\endcsname\relax\def\natexlab#1{#1}\fi
\providecommand{\url}[1]{\href{#1}{#1}}
\providecommand{\dodoi}[1]{doi:~\href{http://doi.org/#1}{\nolinkurl{#1}}}
\providecommand{\doeprint}[1]{\href{http://ascl.net/#1}{\nolinkurl{http://ascl.net/#1}}}
\providecommand{\doarXiv}[1]{\href{https://arxiv.org/abs/#1}{\nolinkurl{https://arxiv.org/abs/#1}}}

\bibitem[{Afonso {et~al.}(2003)Afonso, Albert, Andersen, Ansari, Aubourg,
  Bareyre, Beaulieu, Blanc, Charlot, Couchot, Coutures, Ferlet, Fouqu{\'{e}},
  Glicenstein, Goldman, Gould, Graff, Gros, Haissinski, Hamadache, de~Kat,
  Lasserre, Guillou, Lesquoy, Loup, Magneville, Marquette, Maurice, Maury,
  Milsztajn, Moniez, Palanque-Delabrouille, Perdereau, Pr{\'{e}}vot, Rahal,
  Rich, Spiro, Tisserand, Vidal-Madjar, Vigroux, \& Zylberajch}]{Afonso2003}
Afonso, C., Albert, J.~N., Andersen, J., {et~al.} 2003, Astronomy {\&}
  Astrophysics, 400, 951, \dodoi{10.1051/0004-6361:20030087}

\bibitem[{Alcock {et~al.}(2000)Alcock, Allsman, Alves, Axelrod, Becker,
  Bennett, Cook, Dalal, Drake, Freeman, Geha, Griest, Lehner, Marshall,
  Minniti, Nelson, Peterson, Popowski, Pratt, Quinn, Stubbs, Sutherland,
  Tomaney, Vandehei, \& Welch}]{Alcock2000}
Alcock, C., Allsman, R.~A., Alves, D.~R., {et~al.} 2000, The Astrophysical
  Journal, 542, 281, \dodoi{10.1086/309512}

\bibitem[{Alcock {et~al.}(2001)Alcock, Allsman, Alves, Axelrod, Becker,
  Bennett, Cook, Dalal, Drake, Freeman, Geha, Griest, Lehner, Marshall,
  Minniti, Nelson, Peterson, Popowski, Pratt, Quinn, Stubbs, Sutherland,
  Tomaney, Vandehei, Welch, \& Collaboration)}]{Alcock2001}
---. 2001, The Astrophysical Journal, 550, L169, \dodoi{10.1086/319636}

\bibitem[{Bellm {et~al.}(2019{\natexlab{a}})Bellm, Kulkarni, Graham, Dekany,
  Smith, Riddle, Masci, Helou, Prince, Adams, Barbarino, Barlow, Bauer, Beck,
  Belicki, Biswas, Blagorodnova, Bodewits, Bolin, Brinnel, Brooke, Bue, Bulla,
  Burruss, Cenko, Chang, Connolly, Coughlin, Cromer, Cunningham, De, Delacroix,
  Desai, Duev, Eadie, Farnham, Feeney, Feindt, Flynn, Franckowiak, Frederick,
  Fremling, Gal-Yam, Gezari, Giomi, Goldstein, Golkhou, Goobar, Groom,
  Hacopians, Hale, Henning, Ho, Hover, Howell, Hung, Huppenkothen, Imel, Ip,
  Ivezi{\'{c}}, Jackson, Jones, Juric, Kasliwal, Kaspi, Kaye, Kelley, Kowalski,
  Kramer, Kupfer, Landry, Laher, Lee, Lin, Lin, Lunnan, Giomi, Mahabal, Mao,
  Miller, Monkewitz, Murphy, Ngeow, Nordin, Nugent, Ofek, Patterson, Penprase,
  Porter, Rauch, Rebbapragada, Reiley, Rigault, Rodriguez, van Roestel,
  Rusholme, van Santen, Schulze, Shupe, Singer, Soumagnac, Stein, Surace,
  Sollerman, Szkody, Taddia, Terek, Sistine, van Velzen, Vestrand, Walters,
  Ward, Ye, Yu, Yan, \& Zolkower}]{Bellm2019a}
Bellm, E.~C., Kulkarni, S.~R., Graham, M.~J., {et~al.} 2019{\natexlab{a}},
  Publications of the Astronomical Society of the Pacific, 131, 018002,
  \dodoi{10.1088/1538-3873/aaecbe}

\bibitem[{Bellm {et~al.}(2019{\natexlab{b}})Bellm, Kulkarni, Barlow, Feindt,
  Graham, Goobar, Kupfer, Ngeow, Nugent, Ofek, Prince, Riddle, Walters, \&
  Ye}]{Bellm2019b}
Bellm, E.~C., Kulkarni, S.~R., Barlow, T., {et~al.} 2019{\natexlab{b}},
  Publications of the Astronomical Society of the Pacific, 131, 068003,
  \dodoi{10.1088/1538-3873/ab0c2a}

\bibitem[{Bennett {et~al.}(2002)Bennett, Becker, Quinn, Tomaney, Alcock,
  Allsman, Alves, Axelrod, Calitz, Cook, Drake, Fragile, Freeman, Geha, Griest,
  Johnson, Keller, Laws, Lehner, Marshall, Minniti, Nelson, Peterson, Popowski,
  Pratt, Quinn, Rhie, Stubbs, Sutherland, Vandehei, \& and}]{Bennett2002}
Bennett, D.~P., Becker, A.~C., Quinn, J.~L., {et~al.} 2002, The Astrophysical
  Journal, 579, 639, \dodoi{10.1086/342225}

\bibitem[{Branch {et~al.}(1999)Branch, Coleman, \& Li}]{Branch1999}
Branch, M.~A., Coleman, T.~F., \& Li, Y. 1999, {SIAM} Journal on Scientific
  Computing, 21, 1, \dodoi{10.1137/s1064827595289108}

\bibitem[{Calcino {et~al.}(2018)Calcino, Garc{\'{\i}}a-Bellido, \&
  Davis}]{Calcino2018}
Calcino, J., Garc{\'{\i}}a-Bellido, J., \& Davis, T.~M. 2018, Monthly Notices
  of the Royal Astronomical Society, 479, 2889, \dodoi{10.1093/mnras/sty1368}

\bibitem[{Carr {et~al.}(2016)Carr, K\"{u}hnel, \& Sandstad}]{Carr2016}
Carr, B., K\"{u}hnel, F., \& Sandstad, M. 2016, Physical Review D, 94,
  \dodoi{10.1103/physrevd.94.083504}

\bibitem[{{Di Stefano} \& Esin(1995)}]{DiStefano1995}
{Di Stefano}, R., \& Esin, A.~A. 1995, The Astrophysical Journal, 448,
  \dodoi{10.1086/309588}

\bibitem[{Durbin \& Watson(1971)}]{Durbin1971}
Durbin, J., \& Watson, G.~S. 1971, Biometrika, 58, 1, \dodoi{10.2307/2334313}

\bibitem[{Einstein(1936)}]{Einstein1936LENS-LIKEFIELD}
Einstein, A. 1936, Science (New York, N.Y.), 84, 506,
  \dodoi{10.1126/science.84.2188.506}

\bibitem[{{Gezer} {et~al.}(2022){Gezer}, {Wyrzykowski}, {Zieli{\'n}ski},
  {Marton}, {Kruszy{\'n}ska}, {Rybicki}, {Ihanec}, {Jab{\l}o{\'n}ska}, \&
  {Zi{\'o}{\l}kowska}}]{Gezer:2022}
{Gezer}, I., {Wyrzykowski}, {\L}., {Zieli{\'n}ski}, P., {et~al.} 2022, arXiv
  e-prints, arXiv:2201.12209.
\newblock \doarXiv{2201.12209}

\bibitem[{{Godines} {et~al.}(2019){Godines}, {Bachelet}, {Narayan}, \&
  {Street}}]{Godines:2019}
{Godines}, D., {Bachelet}, E., {Narayan}, G., \& {Street}, R.~A. 2019,
  Astronomy and Computing, 28, 100298, \dodoi{10.1016/j.ascom.2019.100298}

\bibitem[{{Golovich} {et~al.}(2022){Golovich}, {Dawson}, {Bartoli{\'c}}, {Lam},
  {Lu}, {Medford}, {Schneider}, {Chapline}, {Schlafly}, {Drlica-Wagner}, \&
  {Pruett}}]{Golovich2022}
{Golovich}, N., {Dawson}, W., {Bartoli{\'c}}, F., {et~al.} 2022, \apjs, 260, 2,
  \dodoi{10.3847/1538-4365/ac5969}

\bibitem[{Gould(1992)}]{Gould1992}
Gould, A. 1992, The Astrophysical Journal, 392, 442, \dodoi{10.1086/171443}

\bibitem[{Graham {et~al.}(2019)Graham, Kulkarni, Bellm, Adams, Barbarino,
  Blagorodnova, Bodewits, Bolin, Brady, Cenko, Chang, Coughlin, De, Eadie,
  Farnham, Feindt, Franckowiak, Fremling, Gezari, Ghosh, Goldstein, Golkhou,
  Goobar, Ho, Huppenkothen, Ivezi{\'{c}}, Jones, Juric, Kaplan, Kasliwal,
  Kelley, Kupfer, Lee, Lin, Lunnan, Mahabal, Miller, Ngeow, Nugent, Ofek,
  Prince, Rauch, van Roestel, Schulze, Singer, Sollerman, Taddia, Yan, Ye, Yu,
  Barlow, Bauer, Beck, Belicki, Biswas, Brinnel, Brooke, Bue, Bulla, Burruss,
  Connolly, Cromer, Cunningham, Dekany, Delacroix, Desai, Duev, Feeney, Flynn,
  Frederick, Gal-Yam, Giomi, Groom, Hacopians, Hale, Helou, Henning, Hover,
  Hillenbrand, Howell, Hung, Imel, Ip, Jackson, Kaspi, Kaye, Kowalski, Kramer,
  Kuhn, Landry, Laher, Mao, Masci, Monkewitz, Murphy, Nordin, Patterson,
  Penprase, Porter, Rebbapragada, Reiley, Riddle, Rigault, Rodriguez, Rusholme,
  van Santen, Shupe, Smith, Soumagnac, Stein, Surace, Szkody, Terek, Sistine,
  van Velzen, Vestrand, Walters, Ward, Zhang, \& Zolkower}]{Graham2019}
Graham, M.~J., Kulkarni, S.~R., Bellm, E.~C., {et~al.} 2019, Publications of
  the Astronomical Society of the Pacific, 131, 078001,
  \dodoi{10.1088/1538-3873/ab006c}

\bibitem[{Kim {et~al.}(2018{\natexlab{a}})Kim, Kim, Hwang, Albrow, Chung,
  Gould, Han, Jung, Ryu, Shin, Yee, Zhu, Cha, Kim, Lee, Lee, Lee, Park, \&
  and}]{Kim_2018}
Kim, D.-J., Kim, H.-W., Hwang, K.-H., {et~al.} 2018{\natexlab{a}}, The
  Astronomical Journal, 155, 76, \dodoi{10.3847/1538-3881/aaa47b}

\bibitem[{Kim {et~al.}(2018{\natexlab{b}})Kim, Kim, Hwang, Albrow, Chung,
  Gould, Han, Jung, Ryu, Shin, Yee, Zhu, Cha, Kim, Lee, Lee, Lee, Park, \&
  and}]{Kim2018}
---. 2018{\natexlab{b}}, The Astronomical Journal, 155, 76,
  \dodoi{10.3847/1538-3881/aaa47b}

\bibitem[{Lam {et~al.}(2020)Lam, Lu, Hosek, Dawson, \& Golovich}]{Lam2020}
Lam, C.~Y., Lu, J.~R., Hosek, M.~W., Dawson, W.~A., \& Golovich, N.~R. 2020,
  The Astrophysical Journal, 889, 31, \dodoi{10.3847/1538-4357/ab5fd3}

\bibitem[{{Lam} {et~al.}(2022){Lam}, {Lu}, {Udalski}, {Bond}, {Bennett},
  {Skowron}, {Mr{\'o}z}, {Poleski}, {Sumi}, {Szyma{\'n}ski}, {Koz{\l}owski},
  {Pietrukowicz}, {Soszy{\'n}ski}, {Ulaczyk}, {Wyrzykowski}, {Miyazaki},
  {Suzuki}, {Koshimoto}, {Rattenbury}, {Hosek}, {Abe}, {Barry}, {Bhattacharya},
  {Fukui}, {Fujii}, {Hirao}, {Itow}, {Kirikawa}, {Kondo}, {Matsubara},
  {Matsumoto}, {Muraki}, {Olmschenk}, {Ranc}, {Okamura}, {Satoh}, {Silva},
  {Toda}, {Tristram}, {Vandorou}, {Yama}, {Abrams}, {Agarwal}, {Rose}, \&
  {Terry}}]{Lam2022}
{Lam}, C.~Y., {Lu}, J.~R., {Udalski}, A., {et~al.} 2022, \apjl, 933, L23,
  \dodoi{10.3847/2041-8213/ac7442}

\bibitem[{Lu {et~al.}(2016)Lu, Sinukoff, Ofek, Udalski, \& Kozlowski}]{Lu2016}
Lu, J.~R., Sinukoff, E., Ofek, E.~O., Udalski, A., \& Kozlowski, S. 2016, The
  Astrophysical Journal, 830, 41, \dodoi{10.3847/0004-637x/830/1/41}

\bibitem[{{Lyke} {et~al.}(2020){Lyke}, {Higley}, {McLane}, {Schurhammer},
  {Myers}, {Ross}, {Dawson}, {Chabanier}, {Martini}, {Busca}, {Mas des
  Bourboux}, {Salvato}, {Streblyanska}, {Zarrouk}, {Burtin}, {Anderson},
  {Bautista}, {Bizyaev}, {Brandt}, {Brinkmann}, {Brownstein}, {Comparat},
  {Green}, {de la Macorra}, {Mu{\~n}oz Guti{\'e}rrez}, {Hou}, {Newman},
  {Palanque-Delabrouille}, {P{\^a}ris}, {Percival}, {Petitjean}, {Rich},
  {Rossi}, {Schneider}, {Smith}, {Vivek}, \& {Weaver}}]{Lyke:2020}
{Lyke}, B.~W., {Higley}, A.~N., {McLane}, J.~N., {et~al.} 2020, \apjs, 250, 8,
  \dodoi{10.3847/1538-4365/aba623}

\bibitem[{Masci {et~al.}(2018)Masci, Laher, Rusholme, Shupe, Groom, Surace,
  Jackson, Monkewitz, Beck, Flynn, Terek, Landry, Hacopians, Desai, Howell,
  Brooke, Imel, Wachter, Ye, Lin, Cenko, Cunningham, Rebbapragada, Bue, Miller,
  Mahabal, Bellm, Patterson, Juri{\'{c}}, Golkhou, Ofek, Walters, Graham,
  Kasliwal, Dekany, Kupfer, Burdge, Cannella, Barlow, Sistine, Giomi, Fremling,
  Blagorodnova, Levitan, Riddle, Smith, Helou, Prince, \& Kulkarni}]{Masci2018}
Masci, F.~J., Laher, R.~R., Rusholme, B., {et~al.} 2018, Publications of the
  Astronomical Society of the Pacific, 131, 018003,
  \dodoi{10.1088/1538-3873/aae8ac}

\bibitem[{Medford(2021{\natexlab{a}})}]{Medford2021}
Medford, M.~S. 2021{\natexlab{a}}, PhD thesis, University of California,
  Berkeley

\bibitem[{Medford(2021{\natexlab{b}})}]{zort}
---. 2021{\natexlab{b}}, zort: ZTF Object Reader Tool
  \url{https://github.com/michaelmedford/zort},  Zenodo,
  \dodoi{10.5281/ZENODO.4678708}

\bibitem[{Medford(2021{\natexlab{c}})}]{puzle_microlensing_catalogs}
---. 2021{\natexlab{c}}, PUZLE Microlensing Catalogs
  \url{https://doi.org/10.5281/zenodo.7277537},  Zenodo,
  \dodoi{10.5281/zenodo.7277537}

\bibitem[{Medford {et~al.}(2020)Medford, Lu, Dawson, Lam, Golovich, Schlafly,
  \& Nugent}]{Medford2020b}
Medford, M.~S., Lu, J.~R., Dawson, W.~A., {et~al.} 2020, The Astrophysical
  Journal, 897, 144, \dodoi{10.3847/1538-4357/ab9a4f}

\bibitem[{{Mr{\'o}z}(2020)}]{Mroz:2020}
{Mr{\'o}z}, P. 2020, \actaa, 70, 169, \dodoi{10.32023/0001-5237/70.3.1}

\bibitem[{{Mr{\'o}z} {et~al.}(2022){Mr{\'o}z}, {Udalski}, \&
  {Gould}}]{Mroz2022}
{Mr{\'o}z}, P., {Udalski}, A., \& {Gould}, A. 2022, \apjl, 937, L24,
  \dodoi{10.3847/2041-8213/ac90bb}

\bibitem[{Mr{\'{o}}z {et~al.}(2019)Mr{\'{o}}z, Udalski, Skowron,
  Szyma{\'{n}}ski, Soszy{\'{n}}ski, Wyrzykowski, Pietrukowicz, Koz{\l}owski,
  Poleski, Ulaczyk, Rybicki, \& Iwanek}]{Mrz2019}
Mr{\'{o}}z, P., Udalski, A., Skowron, J., {et~al.} 2019, The Astrophysical
  Journal Supplement Series, 244, 29, \dodoi{10.3847/1538-4365/ab426b}

\bibitem[{Mr{\'{o}}z {et~al.}(2020{\natexlab{a}})Mr{\'{o}}z, Udalski,
  Szyma{\'{n}}ski, Soszy{\'{n}}ski, Pietrukowicz, Koz{\l}owski, Skowron,
  Poleski, Ulaczyk, Gromadzki, Rybicki, Iwanek, \& Wrona}]{Mroz2020A}
Mr{\'{o}}z, P., Udalski, A., Szyma{\'{n}}ski, M.~K., {et~al.}
  2020{\natexlab{a}}, The Astrophysical Journal Supplement Series, 249, 16,
  \dodoi{10.3847/1538-4365/ab9366}

\bibitem[{Mr{\'{o}}z {et~al.}(2020{\natexlab{b}})Mr{\'{o}}z, Street, Bachelet,
  Ofek, Bellm, Dekany, Duev, Gal-Yam, Graham, Masci, Porter, Rusholme, Smith,
  Soumagnac, \& Zolkower}]{Mroz2020B}
Mr{\'{o}}z, P., Street, R.~A., Bachelet, E., {et~al.} 2020{\natexlab{b}},
  Research Notes of the {AAS}, 4, 13, \dodoi{10.3847/2515-5172/ab7021}

\bibitem[{Möller {et~al.}(2020)Möller, Peloton, Ishida, Arnault, Bachelet,
  Blaineau, Boutigny, Chauhan, Gangler, Hernandez, Hrivnac, Leoni, Leroy,
  Moniez, Pateyron, Ramparison, Turpin, Ansari, Jr, Bajat, Biswas, Boucaud,
  Bregeon, Campagne, Cohen-Tanugi, Coleiro, Dornic, Fouchez, Godet, Gris,
  Karpov, Gomez-Moran, Neveu, Plaszczynski, Savchenko, \& Webb}]{Moller_2020}
Möller, A., Peloton, J., Ishida, E. E.~O., {et~al.} 2020, Monthly Notices of
  the Royal Astronomical Society, 501, 3272, \dodoi{10.1093/mnras/staa3602}

\bibitem[{Navarro {et~al.}(2017)Navarro, Minniti, \& Ramos}]{Navarro2017}
Navarro, M.~G., Minniti, D., \& Ramos, R.~C. 2017, The Astrophysical Journal,
  851, L13, \dodoi{10.3847/2041-8213/aa9b29}

\bibitem[{Niikura {et~al.}(2019)Niikura, Takada, Yasuda, Lupton, Sumi, More,
  Kurita, Sugiyama, More, Oguri, \& Chiba}]{Niikura_2019}
Niikura, H., Takada, M., Yasuda, N., {et~al.} 2019, Nature Astronomy, 3, 524,
  \dodoi{10.1038/s41550-019-0723-1}

\bibitem[{Novati {et~al.}(2014)Novati, Bozza, Bruni, Dall'Ora, Paolis, Dominik,
  Gualandi, Ingrosso, Jetzer, Mancini, Nucita, Safonova, Scarpetta, Sereno,
  Strafella, Subramaniam, \& Gould}]{Calchi_Novati_2014}
Novati, S.~C., Bozza, V., Bruni, I., {et~al.} 2014, The Astrophysical Journal,
  783, 86, \dodoi{10.1088/0004-637x/783/2/86}

\bibitem[{Paczy{\'{n}}ski(1986)}]{Paczynski1986}
Paczy{\'{n}}ski, B. 1986, The Astrophysical Journal, 304, 1,
  \dodoi{10.1086/164140}

\bibitem[{Powell(1964)}]{Powell1964}
Powell, M. J.~D. 1964, The Computer Journal, 7, 155,
  \dodoi{10.1093/comjnl/7.2.155}

\bibitem[{Price-Whelan {et~al.}(2014)Price-Whelan, Ag\"{u}eros, Fournier,
  Street, Ofek, Covey, Levitan, Laher, Sesar, \& Surace}]{PriceWhelan2014}
Price-Whelan, A.~M., Ag\"{u}eros, M.~A., Fournier, A.~P., {et~al.} 2014, The
  Astrophysical Journal, 781, 35, \dodoi{10.1088/0004-637x/781/1/35}

\bibitem[{{Prince} \& {Zwicky Transient Facility Project
  Team}(2018)}]{Prince_2018}
{Prince}, T., \& {Zwicky Transient Facility Project Team}. 2018, in American
  Astronomical Society Meeting Abstracts, Vol. 231, American Astronomical
  Society Meeting Abstracts \#231, 348.18

\bibitem[{{Pye} \& {McHardy}(1983)}]{Pye:1983}
{Pye}, J.~P., \& {McHardy}, I.~M. 1983, \mnras, 205, 875,
  \dodoi{10.1093/mnras/205.3.875}

\bibitem[{Rahal {et~al.}(2009)Rahal, Afonso, Albert, Andersen, Ansari, Aubourg,
  Bareyre, Beaulieu, Charlot, Couchot, Coutures, Derue, Ferlet, Fouqu{\'{e}},
  Glicenstein, Goldman, Gould, Graff, Gros, Haïssinski, Hamadache, de~Kat,
  Lesquoy, Loup, Guillou, Magneville, Mansoux, Marquette, Maurice, Maury,
  Milsztajn, Moniez, Palanque-Delabrouille, Perdereau, Rahvar, Rich, Spiro,
  Tisserand, \& and}]{Rahal2009}
Rahal, Y.~R., Afonso, C., Albert, J.-N., {et~al.} 2009, Astronomy {\&}
  Astrophysics, 500, 1027, \dodoi{10.1051/0004-6361/200811515}

\bibitem[{Refsdal \& Bondi(1964)}]{Refsdal1964}
Refsdal, S., \& Bondi, H. 1964, Monthly Notices of the Royal Astronomical
  Society, 128, 295, \dodoi{10.1093/mnras/128.4.295}

\bibitem[{{Rodriguez} {et~al.}(2022){Rodriguez}, {Mr{\'o}z}, {Kulkarni},
  {Andreoni}, {Bellm}, {Dekany}, {Drake}, {Duev}, {Graham}, {Masci}, {Prince},
  {Riddle}, \& {Shupe}}]{Rodriguez:2022}
{Rodriguez}, A.~C., {Mr{\'o}z}, P., {Kulkarni}, S.~R., {et~al.} 2022, \apj,
  927, 150, \dodoi{10.3847/1538-4357/ac51cc}

\bibitem[{{Sahu} {et~al.}(2022){Sahu}, {Anderson}, {Casertano}, {Bond},
  {Udalski}, {Dominik}, {Calamida}, {Bellini}, {Brown}, {Rejkuba}, {Bajaj},
  {Kains}, {Ferguson}, {Fryer}, {Yock}, {Mr{\'o}z}, {Koz{\l}owski},
  {Pietrukowicz}, {Poleski}, {Skowron}, {Soszy{\'n}ski}, {Szyma{\'n}ski},
  {Ulaczyk}, {Wyrzykowski}, {Barry}, {Bennett}, {Bond}, {Hirao}, {Silva},
  {Kondo}, {Koshimoto}, {Ranc}, {Rattenbury}, {Sumi}, {Suzuki}, {Tristram},
  {Vandorou}, {Beaulieu}, {Marquette}, {Cole}, {Fouqu{\'e}}, {Hill}, {Dieters},
  {Coutures}, {Dominis-Prester}, {Bennett}, {Bachelet}, {Menzies}, {Albrow},
  {Pollard}, {Gould}, {Yee}, {Allen}, {Almeida}, {Christie}, {Drummond},
  {Gal-Yam}, {Gorbikov}, {Jablonski}, {Lee}, {Maoz}, {Manulis}, {McCormick},
  {Natusch}, {Pogge}, {Shvartzvald}, {J{\o}rgensen}, {Alsubai}, {Andersen},
  {Bozza}, {Novati}, {Burgdorf}, {Hinse}, {Hundertmark}, {Husser}, {Kerins},
  {Longa-Pe{\~n}a}, {Mancini}, {Penny}, {Rahvar}, {Ricci}, {Sajadian},
  {Skottfelt}, {Snodgrass}, {Southworth}, {Tregloan-Reed}, {Wambsganss},
  {Wertz}, {Tsapras}, {Street}, {Bramich}, {Horne}, {Steele}, \& {RoboNet
  Collaboration}}]{Sahu2022}
{Sahu}, K.~C., {Anderson}, J., {Casertano}, S., {et~al.} 2022, \apj, 933, 83,
  \dodoi{10.3847/1538-4357/ac739e}

\bibitem[{{Sajadian} \& {Poleski}(2019)}]{Sajadian_2019}
{Sajadian}, S., \& {Poleski}, R. 2019, \apj, 871, 205,
  \dodoi{10.3847/1538-4357/aafa1d}

\bibitem[{{S{\'a}nchez-S{\'a}ez} {et~al.}(2021){S{\'a}nchez-S{\'a}ez}, {Reyes},
  {Valenzuela}, {F{\"o}rster}, {Eyheramendy}, {Elorrieta}, {Bauer},
  {Cabrera-Vives}, {Est{\'e}vez}, {Catelan}, {Pignata}, {Huijse}, {De Cicco},
  {Ar{\'e}valo}, {Carrasco-Davis}, {Abril}, {Kurtev}, {Borissova}, {Arredondo},
  {Castillo-Navarrete}, {Rodriguez}, {Ruz-Mieres}, {Moya},
  {Sabatini-Gacit{\'u}a}, {Sep{\'u}lveda-Cobo}, \&
  {Camacho-I{\~n}iguez}}]{Sanchez2021}
{S{\'a}nchez-S{\'a}ez}, P., {Reyes}, I., {Valenzuela}, C., {et~al.} 2021, \aj,
  161, 141, \dodoi{10.3847/1538-3881/abd5c1}

\bibitem[{Sumi {et~al.}(2013)Sumi, Bennett, Bond, Abe, Botzler, Fukui,
  Furusawa, Itow, Ling, Masuda, Matsubara, Muraki, Ohnishi, Rattenbury, Saito,
  Sullivan, Suzuki, Sweatman, Tristram, Wada, \& and}]{Sumi2013}
Sumi, T., Bennett, D.~P., Bond, I.~A., {et~al.} 2013, The Astrophysical
  Journal, 778, 150, \dodoi{10.1088/0004-637x/778/2/150}

\bibitem[{Tachibana \& Miller(2018)}]{Tachibana2018}
Tachibana, Y., \& Miller, A.~A. 2018, Publications of the Astronomical Society
  of the Pacific, 130, 128001, \dodoi{10.1088/1538-3873/aae3d9}

\bibitem[{Tisserand {et~al.}(2007)Tisserand, Guillou, Afonso, Albert, Andersen,
  Ansari, Aubourg, Bareyre, Beaulieu, Charlot, Coutures, Ferlet, Fouqu{\'{e}},
  Glicenstein, Goldman, Gould, Graff, Gros, Haissinski, Hamadache, de~Kat,
  Lasserre, Lesquoy, Loup, Magneville, Marquette, Maurice, Maury, Milsztajn,
  Moniez, Palanque-Delabrouille, Perdereau, Rahal, Rich, Spiro, Vidal-Madjar,
  \& and}]{Tisserand2007}
Tisserand, P., Guillou, L.~L., Afonso, C., {et~al.} 2007, Astronomy {\&}
  Astrophysics, 469, 387, \dodoi{10.1051/0004-6361:20066017}

\bibitem[{Udalski {et~al.}(2015)Udalski, Szymański, \&
  Szymański}]{Udalski2015}
Udalski, A., Szymański, M., \& Szymański, G. 2015, Acta Astronomica, 65

\bibitem[{Virtanen {et~al.}(2020)Virtanen, Gommers, Oliphant, Haberland, Reddy,
  Cournapeau, Burovski, Peterson, Weckesser, Bright, {van der Walt}, Brett,
  Wilson, Millman, Mayorov, Nelson, Jones, Kern, Larson, Carey, Polat, Feng,
  Moore, {VanderPlas}, Laxalde, Perktold, Cimrman, Henriksen, Quintero, Harris,
  Archibald, Ribeiro, Pedregosa, {van Mulbregt}, \& {SciPy 1.0
  Contributors}}]{scipy}
Virtanen, P., Gommers, R., Oliphant, T.~E., {et~al.} 2020, Nature Methods, 17,
  261, \dodoi{10.1038/s41592-019-0686-2}

\bibitem[{von Neumann {et~al.}(1941)von Neumann, Kent, Bellinson, \&
  Hart}]{vonNeumann1941}
von Neumann, J., Kent, R.~H., Bellinson, H.~R., \& Hart, B.~I. 1941, The Annals
  of Mathematical Statistics, 12, 153 , \dodoi{10.1214/aoms/1177731746}

\bibitem[{Wyrzykowski {et~al.}(2011{\natexlab{a}})Wyrzykowski, Koz{\l}owski,
  Skowron, Udalski, Szyma{\'{n}}ski, Kubiak, Pietrzy{\'{n}}ski,
  Soszy{\'{n}}ski, Szewczyk, Ulaczyk, \& Poleski}]{Wyrzykowski2011a}
Wyrzykowski, {\L}., Koz{\l}owski, S., Skowron, J., {et~al.} 2011{\natexlab{a}},
  Monthly Notices of the Royal Astronomical Society, 413, 493,
  \dodoi{10.1111/j.1365-2966.2010.18150.x}

\bibitem[{Wyrzykowski {et~al.}(2011{\natexlab{b}})Wyrzykowski, Skowron,
  Koz{\l}owski, Udalski, Szyma{\'{n}}ski, Kubiak, Pietrzy{\'{n}}ski,
  Soszy{\'{n}}ski, Szewczyk, Ulaczyk, Poleski, \& Tisserand}]{Wyrzykowski2011b}
Wyrzykowski, L., Skowron, J., Koz{\l}owski, S., {et~al.} 2011{\natexlab{b}},
  Monthly Notices of the Royal Astronomical Society, 416, 2949,
  \dodoi{10.1111/j.1365-2966.2011.19243.x}

\bibitem[{Wyrzykowski {et~al.}(2015)Wyrzykowski, Rynkiewicz, Skowron,
  Koz{\l}owski, Udalski, Szyma{\'{n}}ski, Kubiak, Soszy{\'{n}}ski,
  Pietrzy{\'{n}}ski, Poleski, Pietrukowicz, \& Pawlak}]{Wyrzykowski2015}
Wyrzykowski, {\L}., Rynkiewicz, A.~E., Skowron, J., {et~al.} 2015, The
  Astrophysical Journal Supplement Series, 216, 12,
  \dodoi{10.1088/0067-0049/216/1/12}

\bibitem[{Wyrzykowski {et~al.}(2016)Wyrzykowski, Kostrzewa-Rutkowska, Skowron,
  Rybicki, Mr{\'{o}}z, Koz{\l}owski, Udalski, Szyma{\'{n}}ski,
  Pietrzy{\'{n}}ski, Soszy{\'{n}}ski, Ulaczyk, Pietrukowicz, Poleski, Pawlak,
  I{\l}kiewicz, \& Rattenbury}]{Wyrzykowski2016}
Wyrzykowski, {\L}., Kostrzewa-Rutkowska, Z., Skowron, J., {et~al.} 2016,
  Monthly Notices of the Royal Astronomical Society, 458, 3012,
  \dodoi{10.1093/mnras/stw426}

\bibitem[{{Zhang} {et~al.}(2021){Zhang}, {Bloom}, {Gaudi}, {Lanusse}, {Lam}, \&
  {Lu}}]{Zhang_lfi:2021}
{Zhang}, K., {Bloom}, J.~S., {Gaudi}, B.~S., {et~al.} 2021, \aj, 161, 262,
  \dodoi{10.3847/1538-3881/abf42e}

\end{thebibliography}

\appendix
\section{Comparison to Rodriguez et al. 2022 and Mr{\'{o}}z et al. 2020b}
\label{Appendix: lit comparison}

\edit{\cite{Rodriguez:2022} (hereafter referred to as \citetalias{Rodriguez:2022}) also searched the ZTF data through DR5 for microlensing events and found 60 events. However, only 28 objects in our sample overlapped with theirs, so we made 32 new detections. Of the 32 objects we found that \citetalias{Rodriguez:2022} did not, 12 were outside of the $|b| < 20^{\circ}$ area they searched. \cite{Mroz2020B} (hereafter \citetalias{Mroz2020B}) searched through DR1 and found 30 events, 12 of which overlapped in our pipeline.}

\edit{Table \ref{tab:pipeline_cuts_lit} summarizes where the 32 and 28 objects \citetalias{Rodriguez:2022} and \citetalias{Mroz2020B}, respectively, identified that were cut in the \texttt{PUZLE} pipeline. Overlapping events refers to events cut that were in both \citetalias{Rodriguez:2022} and \citetalias{Mroz2020B}. 5 and 4 events, respectively (2 of which overlap) were cut by the PanSTARRS1 star-determination score which was necessary in the \texttt{PUZLE} pipeline since we considered objects outside the galactic plane. 6 and 5 (3 of which overlap) were not able to be fit by the 4-parameter model fit, and 9 and 2 were cut by not passing the $\eta_{\rm residual}$ cut. These may be because the \texttt{PUZLE} pipeline always fits the band with the most data whereas \citetalias{Rodriguez:2022} fits the r-band. Some visually inspected events appear to have outliers in the g band or some variability in the baseline which may have caused the fit or the $\eta_{\rm residual}$ cut to fail. 2 and 3 (1 of which overlaps) events either had the peak start before the survey began or after it ended ($t_0 \pm t_E$ in the data). The equivalent cut in \citetalias{Rodriguez:2022} uses $t_0 \pm 0.3t_E$ which may cause the difference. 1 event in \citetalias{Rodriguez:2022} did not pass the $\chi^2_{red}$ cut. 4 events in \citetalias{Rodriguez:2022} did not pass the fractional $t_E$ error cut for which \citetalias{Rodriguez:2022} does not have an equivalent cut. 1 event in both \citetalias{Rodriguez:2022}  and \citetalias{Mroz2020B} did not pass the $|u_0|$ cut. 2 and 1 events (1 of which overlaps) did not have enough baseline data which is important for characterization. 2 events (overlapped in both catalogs) were determined not to be microlensing in the by-eye characterization. 226524\_5543 is missing data in its peak, but there is some rise and fall so it was put in the ``possible microlensing" category, and 214154\_4399 has characteristics of a SNe including a sharp rise and a slower fall and the r-band data falling above the model.}

\edit{One other object of note is ZTF18absrqlr. Two ZTF data brokers, Fink \citep{Moller_2020} and ALeRCE \citep{Sanchez2021}, mark its RA and Dec as $283.84059^{\circ}$, $5.74372^{\circ}$, but it is $307.149376^{\circ}$, $22.830478^{\circ}$ in \citetalias{Rodriguez:2022} and \citetalias{Mroz2020B}. The coordinates $307.149376^{\circ}$, $22.830478^{\circ}$ are duplicated in \citetalias{Mroz2020B} between ZTF18abnbmsr and ZTF18absrqlr. We believe that \citetalias{Rodriguez:2022} actually discovered ZTF18abnbmsr, as we did, and that the ZTF ID is a typo. In Table \ref{tab:pipeline_cuts_lit}, we treated it as having coordinates $283.84059^{\circ}$, $5.74372^{\circ}$ in \citetalias{Mroz2020B} and $307.149376^{\circ}$, $22.830478^{\circ}$ in \citetalias{Rodriguez:2022}.}

\begin{table*}[b]
    \centering
        \begin{tabular}{r|c|c}
             Cut Performed & R22 Cands  & M20 Cands  \\
             \textcolor{white}{whitespace}& Remaining & Remaining \\
             \hline 
             Cands in Paper & 60 & 30 \\
             PS1-PSC $\geq 0.645$ & 55 & 26 \\
             Successful 4-parameter model fit & 49 & 21 \\
             $\eta_\text{residual}\ \geq \eta\ * 3.82 - 0.077$ & 40 & 19 \\
             7-parameter fit: $t_0 - t_E \geq 58194$  & 39 & 16 \\
             $\chi_{red}^2 \leq 3$ & 38 & 16 \\
             $t_0 + t_E \leq 59243$ & 37 & 16 \\
             $\sigma_{t_E} / t_E \leq 0.20$ & 33 & 16 \\
             $|u_0| \leq 1.0$ & 32 & 15 \\
             $4t_E$ baseline outside of $t_0 \pm 2t_E$ & 30 & 14 \\
             Manually assigned clear microlensing label & 28 & 12 \\
        \end{tabular}
        \caption{\edit{Version of Table \ref{tab:pipeline_cuts} with relevant cuts and candidate events from \cite{Rodriguez:2022} and \cite{Mroz2020B} to demonstrate where in the \texttt{PUZLE} pipeline the candidates would have been cut out. We recover 28/60 of \cite{Rodriguez:2022} events and find 1 in the ongoing catalog, and we recover 12/30 of \cite{Mroz2020B} events. See Appendix \ref{Appendix: lit comparison} for a detailed discussion.}}
    \label{tab:pipeline_cuts_lit}
\end{table*}

\pagebreak

\section{Level 6 Events Summary}
\label{Appendix:level 6}
\begin{longtable}[]{r|c|c|c|c|c|c}
    \centering
         puzle ID & $\ell$ (deg) & b (deg) & RA (deg) & Dec (deg) & $m_{\rm base}$ & ZTF ID  \\
         \hline 
         206883\_29398 & 38.01100 & -6.11373 & 290.61721 & 1.70650 & 14.32 $\pm$ 0.00 & ZTF18abaqxrt \\
         226244\_24695 & 133.60050 & -4.54272 & 32.49736 & 56.69314 & 14.91 $\pm$ 0.00 & ZTF21aalljap  \\
         213100\_12775 & 52.73575 & -2.90082 & 294.83615 & 16.14483 & 15.76 $\pm$ 0.00 & ZTF18abibobd  \\
         226146\_9143 & 89.38878 & 12.39568 & 301.31815 & 55.32261 & 16.43 $\pm$ 0.00 & ZTF20abbynqb \\
         199814\_8384 & 14.10087 & 2.38621 & 271.84240 & -15.54738 & 16.76 $\pm$ 0.00 & ZTF19abbwpl \\
         211087\_11025 & 41.40802 & 8.11634 & 279.40461 & 11.20061 & 16.92 $\pm$ 0.00 & ZTF19aaekacq  \\
         213871\_3211 & 66.29863 & -19.29724 & 316.81620 & 18.13759 & 17.03 $\pm$ 0.00 & ZTF19aacmpke  \\
         219115\_17109 & 73.66961 & -4.80719 & 309.03411 & 32.72093 & 17.08 $\pm$ 0.01 & ZTF19aavnrqt  \\
         216020\_13108 & 60.31419 & 1.48227 & 294.73270 & 24.89685 & 17.31 $\pm$ 0.00 & ZTF18abjnrmm  \\
         215639\_28569 & 57.62132 & 4.06236 & 290.83245 & 23.77312 & 17.48 $\pm$ 0.00 & ZTF20abkyuyk  \\
         208826\_2418 & -61.55329 & 69.64089 & 191.28933 & 6.82385 & 17.52 $\pm$ 0.00 & ZTF18acnokyi  \\
         198333\_4452 & -40.86573 & 42.90427 & 205.29821 & -18.39670 & 17.61 $\pm$ 0.01 & ZTF19aarihyl  \\
         221243\_1676 & 85.63918 & -9.07047 & 322.66138 & 38.85054 & 17.65 $\pm$ 0.01 & ZTF20abicdkm  \\
         223737\_470 & 150.23484 & -6.02984 & 55.43845 & 47.60916 & 17.68 $\pm$ 0.00 & - \\
         200772\_6200 & 21.83300 & -9.02418 & 285.98402 & -13.92945 & 17.68 $\pm$ 0.00 & ZTF18abmoxlq  \\
         204396\_25713 & 29.37656 & -2.38778 & 283.36662 & -4.25183 & 17.81 $\pm$ 0.00 & ZTFJ1853.5-0415 \\
         199420\_23799 & 17.46218 & -5.58708 & 280.80561 & -16.29913 &	17.92 $\pm$ 0.01 & ZTF19abqueuo  \\
         229329\_2194 & 143.55312 & 35.15946 & 132.46388 & 70.66061 & 17.96 $\pm$ 0.01 & ZTF19adbtqgz  \\
         215476\_198 & -170.52767 & 5.21321 & 96.61807 & 23.22059 & 18.00 $\pm$ 0.00 & -  \\
         203599\_7926 & 26.58814 & -1.75180 & 281.52435 & -6.44398 & 18.13 $\pm$ 0.01 & ZTF19aargwsl  \\
         198045\_37053 & 12.53102 & -2.35737 & 275.42456 & -19.19178 & 18.22 $\pm$ 0.01 & -  \\
         227712\_653 & 138.73891 & 3.95527 & 48.69432 & 62.34347 & 18.37 $\pm$ 0.01 & ZTF19aabbuqn  \\
         215341\_28211 & 64.60030 & -9.23556 & 307.14936 & 22.83048 & 18.39 $\pm$ 0.00 & ZTF18abnbmsr  \\
         198550\_16292 & 17.54506 & -9.86895 & 284.86976 & -18.10769 & 18.43 $\pm$ 0.01 & ZTF20aawxugf  \\
         207262\_13818 & 34.15426 & 2.63134 & 281.07346 & 2.28667 & 18.53 $\pm$ 0.02 & ZTF19aaimlse  \\
         227808\_6978 & 96.97033 & 45.32348 & 234.62998 & 62.48028 & 18.57 $\pm$ 0.02 & ZTF21aazeazr, ZTF20aaljffg \\
         220863\_6930 & 69.57788 & 10.90194 & 290.01643 & 37.43482 & 18.61 $\pm$ 0.02 & ZTF20aavmhsg  \\
         229865\_2362 & 141.03852 & 30.61447 & 119.55902 & 73.78493 & 18.74 $\pm$ 0.01 & ZTF19adceqzb  \\
          224227\_29874 & 93.23237 & -2.93470 & 324.57735 & 48.47926 & 18.76 $\pm$ 0.00 & ZTF20abvwhlb \\
         204386\_5100 & 28.60496 &-1.06885 & 281.83689 & -4.33811 & 18.88 $\pm$ 0.01 & ZTF19aavndrc \\
         205166\_6377 & 26.80213 & 6.92269 & 273.90054 & -2.25697 & 18.97 $\pm$ 0.01 & ZTF19aaonska \\
         205621\_17625 & 32.19706 & -1.09287 & 283.49719 & -1.15227 & 18.98 $\pm$ 0.02 & ZTF19aaxsdqz \\
         216370\_6986 & 60.09199 & 3.08836 & 293.06192 & 25.48256 & 19.06 $\pm$ 0.01 & ZTF20abmxjsq  \\
         206452\_5009 & 34.18144 & -2.37119 & 285.53940 & 0.03010 & 19.07 $\pm$ 0.01 & ZTFJ1902.2+0001  \\
         213011\_4595 & 29.87173 & 42.93640 & 242.37401 & 16.01123 & 19.28 $\pm$ 0.01 & ZTF19aasbpld  \\
         222741\_5669 & 86.49219 & -3.55857 & 318.26336 & 43.33765 & 19.33 $\pm$ 0.01 & ZTF19abftuld  \\
         225310\_3105 & 81.71923 & 22.43713 & 280.73170 & 52.45385 & 19.37 $\pm$ 0.01 & ZTF19abbyebj  \\
         221496\_7115 & 72.90355 & 8.32557 & 294.84456 & 39.17829 & 19.38 $\pm$ 0.01 & ZTF20abrtvbz  \\
         197737\_7136 & 46.67683 & -68.20361 & 350.26232 & -20.17373 & 19.38 $\pm$ 0.02 &  ZTF20acbdmhy  \\
         205225\_28572 & 31.55493 & -3.71994 & 285.54582 & -2.91912 & 19.43 $\pm$ 0.01 & ZTF19acctqyc \\
         201592\_4317 & 23.84423 & -7.34805 & 285.33834 & -11.40450 & 19.45 $\pm$ 0.02 & ZTF18absjezs  \\
         225706\_1669 & 133.87116 & -6.45141 & 31.94920 & 54.79037 & 19.52 $\pm$ 0.01 & -  \\
         209120\_548 & 120.81612 & -55.66621 & 11.65688 & 7.18577 & 19.62 $\pm$ 0.01 & ZTF18abwamwf \\
         198940\_35883 & 14.27337 & -1.33996 & 275.33980 & -17.17655 & 19.63 $\pm$ 0.03 & -  \\
         210144\_774 & 22.08289 & 40.47883 & 242.13879 & 9.70584 & 19.72 $\pm$ 0.02 & ZTF20aawghfe  \\
         225914\_13362 & 99.04746 & -0.46726 & 329.19298 & 54.09859 & 19.75 $\pm$ 0.01 & ZTF18aayhjoe  \\
         210677\_14481 & 42.48555 & 3.54901 & 284.04933 & 10.11678 & 19.78 $\pm$ 0.01 & ZTFJ1856.2+1007  \\
         204863\_2897 & 34.50146 & -11.49132 & 293.81041 & -3.83897 & 19.84 $\pm$ 0.01 & ZTF20aawijop  \\
         219013\_4366 & 62.12153 & 15.97480 & 280.73453 & 32.87309 & 19.86 $\pm$ 0.01 & ZTF19aamrjmu \\
         206800\_23883 & 29.59880 & 10.33983 & 272.14526 & 1.77291 & 19.96 $\pm$ 0.02 & ZTF20abagzwt \\
         203625\_1402 & 28.59530 & -4.97245 & 285.32605 & -6.11747 & 20.01 $\pm$ 0.03 & -  \\
         201089\_14825 & 17.00608 & 4.44172 & 271.43912 & -12.01451 & 20.18 $\pm$ 0.02 & ZTF18ablrdcc \\
         215636\_7171 & 56.73376 & 4.52214 & 289.93921 & 23.20425 & 20.26 $\pm$ 0.01 & ZTF20acbmtwd  \\
         222375\_11305 & 72.03940 & 17.58765 & 283.26664 & 42.29320 & 20.51 $\pm$ 0.03 & -  \\
         219134\_11822 & 77.28109 & -9.13134 & 315.71335 & 32.81170 & 20.65 $\pm$ 0.05 & ZTF19acbptvn  \\
         201147\_4625 & 20.90054 & -4.85962 & 281.72420 & -12.91335 & 20.70 $\pm$ 0.03 & ZTF19abijroe  \\
         227178\_28049 & 101.10161 & 4.66963 & 326.17312 & 59.37792 & 20.81 $\pm$ 0.02 & ZTF18aaztjyd  \\
         218269\_30320 & 64.29643 & 6.23607 & 292.17602 & 30.66463 & 21.21 $\pm$ 0.03 & ZTFJ1928.7+3039 \\
         229949\_8332 & 104.65664 & 24.98626 & 286.18769 & 73.45821 & 21.25 $\pm$ 0.06 & ZTF20aaymulz  \\
         227822\_8865 & 92.34280 & 34.10588 & 260.41834 & 62.87877 & 21.38 $\pm$ 0.04 & ZTF19abgvmln  \\
    \caption{\edit{Summary of the 60 level 6 candidate events ordered by baseline magnitude. More extensive parameters of the events can be found online at \cite{puzle_microlensing_catalogs}.}}
    \label{tab:events}
\end{longtable}
\end{document}